\documentclass[a4paper,11pt]{article}
\pdfoutput=1

\usepackage{custom_style_file}

\usepackage{graphicx,epsfig,natbib,color,times,bm,amsmath,multirow}
\usepackage[ddmmyyyy,hhmmss]{datetime}

\begin{document}
\title{On the Signal Processing Operations in LIGO signals}

\author[a]{{Akhila Raman}},

\affiliation[a]{University of California at Berkeley}

\emailAdd{akhila.raman@berkeley.edu}

\abstract{
This article analyzes the data for the four gravitational wave (GW) events detected in Hanford(H1) and Livingston(L1) detectors by the LIGO\footnote{The Laser Interferometer Gravitational-Wave Observatory} collaboration and also two more events detected by H1, L1 and Virgo(V1) detectors.  It is shown that GW170608, GW170814, GW151226 and GW170104 are very weak signals whose amplitude does not rise significantly during the GW event, and they are indistinguishable from non-stationary detector noise. 

LIGO software implements cross-correlation funcion(CCF) of H1/L1 signals with the template reference signal, in frequency domain, in a matched filter, using 32 second windows. It is shown that this matched filter misfires with high SNR/CCF peaks, even for very low-amplitude, short bursts of sine wave signals and additive white gaussian noise(AWGN), all the time. It is shown that this erratic behaviour of the matched filter, is due to the error in signal processing operations, such as lack of cyclic prefix necessary to account for circular convolution and error in whitening operations. It is also shown that normalized CCF method implemented in time domain using short windows, does not have false CCF peaks for sine wave and noise bursts.

It is shown that the normalized CCF for GW151226 and GW170104, when correlating H1/L1 and template, is indistinguishable from correlating detector noise and the template. It is also shown that the normalized CCF for GW151226 and GW170104, when correlating H1/L1 and template, is indistinguishable from correlating the template vs bogus chirp templates which are frequency modulated(FM) waveforms which differ significantly from ideal templates. Similar results are shown with LIGO matched filter, which misfires with high Signal to Noise Ratio(SNR) for bogus chirp templates. 

Hence it is argued that H1 vs L1 cross-correlation test is necessary to avoid this wrong classification of detector noise and bogus templates as GW signals.  It is shown that normalized CCF is poor, when correlating H1 and L1, which is indistinguishable from detector noise correlations, for GW151226, GW170104, GW170608 and GW170814 and hence it is suggested that they are further studied as candidates for GW signals. The implications of these results are discussed for GW150914 and GW170817 and the possibility of electro-magnetic interference is discussed. All the results in this paper are demonstrated using modified versions of LIGO's Python scripts\citep{LIGO demos}.\footnote{The specific Python script used to generate Fig.1 to Fig.23 in this manuscript, is mentioned in \citep{LIGO demos}.  }
}

\maketitle


\section{Introduction}
\label{sec:introduction}

\begin{figure}[h!]
\centering

\includegraphics[width=0.32\textwidth]{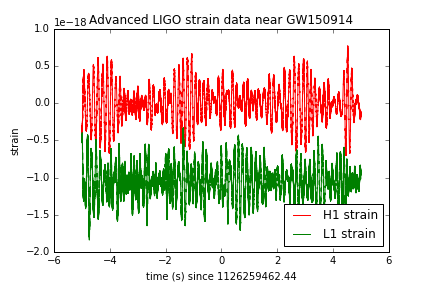}
\includegraphics[width=0.32\textwidth]{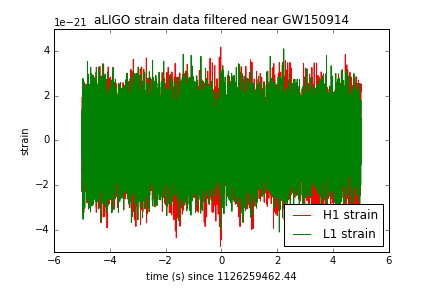}
\includegraphics[width=0.32\textwidth]{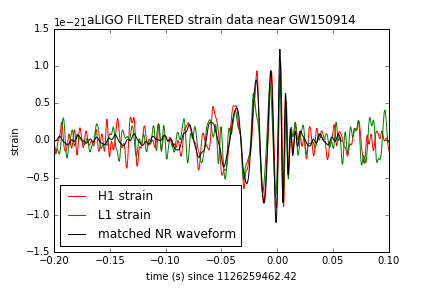}

\caption{GW150914 H1 and L1 strains: Left Panel: Raw strain. Middle Panel: Raw strain filtered in 43-300Hz range. Right Panel: LIGO whitened and filtered strains}
\label{fig:ligo_whitening_0}
\end{figure}

The first \href{https://en.wikipedia.org/wiki/Gravitational_wave}{gravitational wave} GW signal observed in the \href{https://en.wikipedia.org/wiki/LIGO}{LIGO } detector was GW150914 \citep{Ligo1}  which is the strongest of the GW signals observed so far. But, we can see from Fig.\ref{fig:ligo_whitening_0} that, the maximum amplitude of the recovered template\footnote{ Template is the reference signal we hope to detect.} ($1.2*10^{-21}$) shown in the right panel, after LIGO whitening procedure(detailed in Section~\ref{sec:ligo_whitening}) and filtered in 43-300Hz range, is $\frac{1}{400}$th of the amplitude of the raw strain\footnote{Raw strain contains detector noise and the short duration GW event, which may contain the template.} in the left panel ($0.5*10^{-18}$). $85$ percent of the template energy is in the 43-300Hz range. We can also see that, the recovered template for GW150914 in the right panel, is submerged below the maximum amplitude of the filtered strain ($3*10^{-21}$) in the middle panel.

The first \href{https://en.wikipedia.org/wiki/Gravitational_wave}{gravitational wave} GW signal observed in the \href{https://en.wikipedia.org/wiki/LIGO}{LIGO } detector was GW150914 \citep{Ligo1}  which was a relatively strong signal whose amplitude, after whitening and filtering\footnote{LIGO detectors have significant \href{https://losc.ligo.org/s/events/GW150914/P1500238/fig1.png}{impulsive} interference at 60*n Hz and other frequencies, hence they are removed by whitening the signal. Then fourth order Butterworth bandpass filters are used in frequency range 43-300Hz for GW150914, 43-800Hz for GW170104 and GW151226.}, rose significantly, well over detector noise level, during the 0.2 second GW event duration. In comparison, the second signal\citep{Ligo2} \href{https://www.ocf.berkeley.edu/~araman/files/ligo_mf/GW150914/figures/GW151226_H1_L1_strain_whitened_not_superimposed.png}{GW151226}, the third signal\citep{Ligo3} \href{https://www.ocf.berkeley.edu/~araman/files/ligo_mf/GW150914/figures/GW170104_H1_L1_strain_whitened_not_superimposed_2.png}{GW170104}, the fourth signal \citep{Ligo-6} \href{https://www.ocf.berkeley.edu/~araman/files/ligo_tests/GW170608/GW170608_H1_L1_strain_whitened_not_superimposed_1.png}{GW170608}, the fifth signal\citep{Ligo-4} \href{https://www.ocf.berkeley.edu/~araman/files/ligo_tests/GW170814/GW170814_cln_not_superimposed_2.png}{GW170814} and the sixth signal\citep{Ligo-5} \href{https://www.ocf.berkeley.edu/~araman/files/ligo_tests/GW170817/GW170817_cln__not_superimposed_2.png}{GW170817} were very weak signals, which look like noise after whitening and filtering and whose signal amplitude does not rise above the detector noise level during the GW event of duration of less than 1 second (Fig.\ref{fig:GW_amplitude}, Fig.\ref{fig:GW151226_signal}, Fig.\ref{fig:GW170104_signal}, Fig.\ref{fig:GW170814_signal}, Fig.\ref{fig:GW170817_signal} ).

This raises the important question of whether GW151226 and GW170104 could have been caused by bogus chirp templates, non-GW signals from other sources or non-stationary detector noise.  Because we are more often likely to observe weak signals which look like noise and whose amplitude does not rise during assumed GW event, it is of paramount importance that we should not classify non-GW signals or noise as GW events. We must insist on high standards before classifying an observed time series as a GW signal.

The organization of this paper is as follows. In Section~\ref{sec:ligo_mf}, it is shown that LIGO matched filter misfires with high SNR/CCF peaks, even for very low-amplitude, short bursts of sine wave signals and additive white gaussian noise(AWGN), all the time. It is shown that this erratic behaviour of the matched filter, is due to the error in signal processing operations. 

In Section~\ref{sec:bogus_chirp}, bogus chirp templates are simulated, which model the ideal template as a frequency modulated signal and then noise is added to the phase.  It will be shown that the LIGO matched filter misfires with high SNR, even for bogus chirp templates added to the detector noise and is indistinguishable from the SNR observed for ideal template, resulting in false coincidence, for GW151226 and GW170104.

Hence it is argued that H1 vs L1 cross-correlation test is necessary to avoid this wrong classification of bogus templates as GW signals, irrespective of the source of these bogus templates. In Section~\ref{sec:test-3b}, using Normalized Cross Correlation Function(CCF) method, it will be shown that the normalized CCF of H1 vs L1, is indistinguishable from detector noise correlations, for GW151226 and GW170104. 
In Section~\ref{sec:test-3c}, it will be shown that the normalized CCF of H1 vs L1, is indistinguishable from detector noise correlations, for GW170814 and GW170608.

\section{ Errors in Signal Processing Operations in LIGO's Matched Filter}
\label{sec:ligo_mf}

The \textbf{core engine} of the LIGO software for identification of GW signals is the matched filter as described in Eq.1-4 in ~\citep{Ligo4}, which is used in two independent search methods, PYCBC and GSTLAL analysis\footnote{ GSTLAL analysis also uses matched filter search, and  as per Page 7 in  ~\citep{Ligo4}, "the data s(t) and templates h(t) are each whitened in the frequency domain by dividing them by an estimate of the  power spectral density of the detector noise." and also "By the convolution theorem, $\rho(t)$ obtained in this manner is the same as the  $\rho(t)$ obtained by frequency domain filtering in Eq. (1)."[in PYCBC analysis]. These equations are implemented in lines 662-740 in the matched filter section of LIGO's tutorial python script.~\citep{gw151226 tutorial}. }.  For the case of PYCBC analysis, matched filter SNR(MF-SNR) $\rho^{2}(t)$ is given as follows.

\begin{align}
\begin{split}
\rho^{2}(t) = \frac{1}{|\langle h|h \rangle|} |\langle s|h \rangle (t)|^{2} \\
 \langle s|h \rangle (t) = 4 \int_{0}^{\infty} \frac{\widehat{s}(f) \widehat{h}^{*}(f)}{S_n(f)} e^{i 2 \pi f t} df
\end{split}
\label{eqn:1}
\end{align}
where $s(t)$ and $h(t)$ are the strain signal and the template respectively and $\widehat{s}(f)$ and $\widehat{h}^{*}(f)$ are the Fourier Transforms of $s(t)$ and $h^{*}(-t)$ respectively and $S_n(f)$ is the power spectral density of the detector noise. In the time domain, this is equivalent to the convolution of whitened version of $s(t)$ with real $h(-t)$, which is equivalent to the Cross-Correlation Function(CCF) of the whitened strain signal $s(t)$ and the template $h(t)$, with a normalization scale factor as follows. \\
\begin{align}
\begin{split}
\langle h|h \rangle = 4 \int_{0}^{\infty} \frac{\widehat{h}(f) \widehat{h}^{*}(f)}{S_n(f)}  df
\end{split}
\label{eqn:2}
\end{align}

It is noted that the above CCF is \textbf{not} normalized to give MF-SNR =1  for an ideal template correlated with itself. The expression for re-weighted SNR is given in Eq.6. in ~\citep{Ligo4} and the re-weighted SNR threshold of 5 is used to declare a GW event.

\subsection{Signal Processing Operations in LIGO's Matched Filter}
\label{sec:ss_anomaly}

LIGO software implements cross-correlation funcion(CCF) of H1/L1 signals with the template reference signal, in frequency domain, in a matched filter, using 32 second windows. It is shown in the subsections below, that this matched filter misfires with high SNR/CCF peaks, even for very low-amplitude, short bursts of sine wave signals and additive white gaussian noise(AWGN), all the time. It is also shown that normalized CCF method implemented in time domain using short windows, does not have false CCF peaks for sine wave and noise bursts (It may produce false CCF peaks for noise burst rarely, but not all the time).

It is shown here that this erratic behaviour of the matched filter, is due to the error in signal processing operations, such as lack of cyclic prefix necessary to account for circular convolution. LIGO matched filter is supposed to implement Eq.~\ref{eqn:1} and we see that the Fourier transform of the whitened strain signal $s(t)$ is multiplied by the complex conjugate of the Fourier transform of the template $h(t)$ and then the inverse Fourier transform is taken. This is equivalent to the \textbf{linear convolution} of whitened version of $s(t)$ with real $h(-t)$, which is Cross-Correlation Function(CCF) of the whitened strain signal $s(t)$ and the template $h(t)$, which is the desired operation. 

However, in  the \textbf{discrete time} implementation in LIGO python script (\href{https://losc.ligo.org/s/events/GW170104/LOSC_Event_tutorial_GW170104.html#Matched-filtering-to-find-the-signal}{code section}), this is equivalent to \textbf{circular convolution} of $s(t)$ with $h(-t)$ in time domain. Circular convolution for discrete time signals  differs from linear convolution (\href{https://en.wikipedia.org/wiki/Circular_convolution}{here}) and \textbf{cyclic prefix}, which is given by the last N samples of $s(t)$, where $N$ is the length of the template $h(t)$,  should be included, to account for the resulting artifacts (\href{https://en.wikipedia.org/wiki/Cyclic_prefix}{here}). LIGO implementation does not provide a cyclic prefix for the time domain signal $s(t)$. This may be one reason why LIGO matched filter misfires with high SNR, even for very low-amplitude, short bursts of sine wave signals and noise burst, all the time.

\subsection*{LIGO's Matched Filter Misfires for Sine Wave and AWGN}
\label{sec:MF_anomaly}

\begin{figure}[h!]
\centering

\includegraphics[width=0.48\textwidth, height=0.16\textheight]{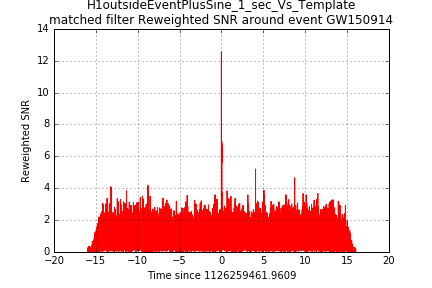}
\includegraphics[width=0.48\textwidth, height=0.16\textheight]{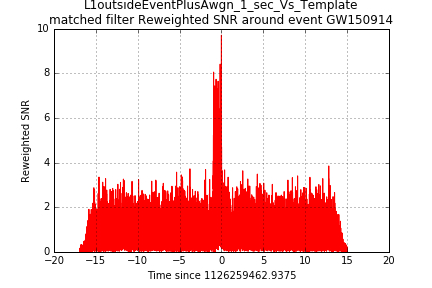}

\includegraphics[width=0.48\textwidth]{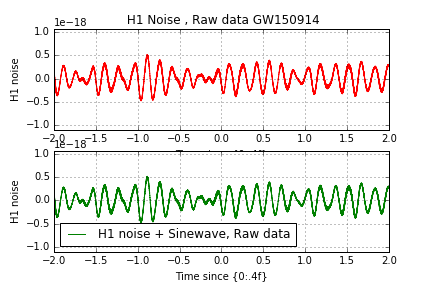}
\includegraphics[width=0.48\textwidth]{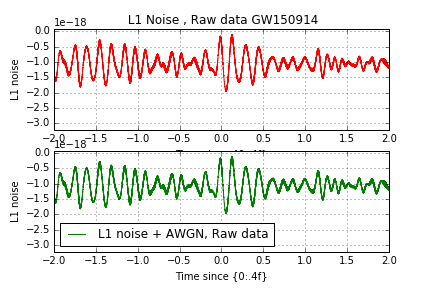}

\includegraphics[width=0.32\textwidth]{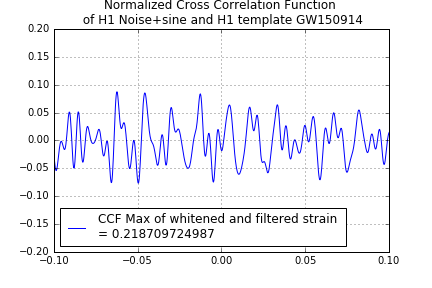}
\includegraphics[width=0.32\textwidth]{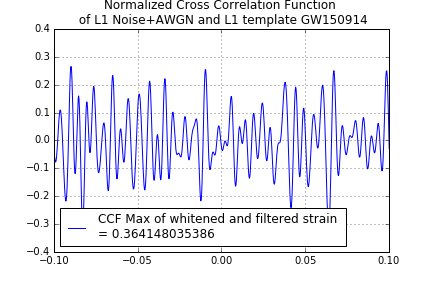}
\includegraphics[width=0.32\textwidth]{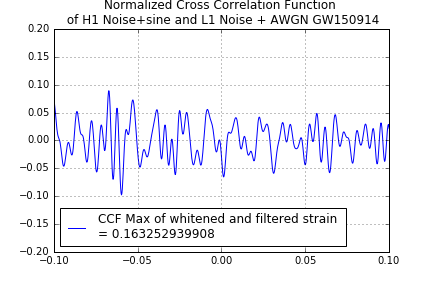}

\caption{Upper Panels: LIGO matched filter Reweighted SNR. Left: Sinewave + H1 detector noise vs template. Right: AWGN + L1 detector noise vs template. Middle panels: Raw unwhitened signals in the time domain. Left: (Sinewave + H1 detector noise) . Right: AWGN + L1 detector noise. Lower Panels: Normalized CCF of whitened and filtered signals. Left: (Sinewave + H1 detector noise) vs template. Middle: (AWGN+ L1 detector noise) vs template. Right: (Sinewave + H1 detector noise) vs (AWGN + L1 detector noise).}
\label{fig:LIGO_matched_filter_sinewave}
\end{figure}

In this section, we will examine whether LIGO's matched filter fires with high SNR, if it is fed with a signal other than the template. Let us test it by replacing the GW150914 signal at H1 with the signal $w_d(t) + a(t)$, where $w_d(t)$ is the raw unwhitened H1 detector noise, obtained from a 32 second block of data which does not contain the GW signal, and $a(t)$ is a sine wave of duration 1 second and frequency $64$ Hz, scaled so that the standard deviation of $a(t)$ is at least 100 times lower than standard deviation of $w_d(t)$ and has an exponential decay. Then we correlate $w_d(t) + a(t)$ with the GW150914 template $h(t)$ in the LIGO matched filter. 

Similarly, we replace the GW150914 signal at L1 with the signal $w_d(t) + b(t)$, where $w_d(t)$ is the raw L1 detector noise, obtained from a 32 second block of data which does not contain the GW signal, and $b(t)$ is AWGN of duration 1 second, scaled so that the standard deviation of $b(t)$ is at least 500 times lower than standard deviation of $w_d(t)$. Then we correlate $w_d(t) + b(t)$ with the GW150914 template $h(t)$ in the LIGO matched filter.

Fig.~\ref{fig:LIGO_matched_filter_sinewave} shows the results, and we can see from upper left panel that, even a low-amplitude sine wave added to the detector noise can produce high reweighted SNR in LIGO's matched filter, well higher than the specified threshold of $5$ . This result is not unique to any specific sine wave frequency, but similar results are observed for other frequencies as well and also for GW151226 and GW170104. Similarly, we can see from the upper right panel that, even low-amplitude white gaussian noise added to the detector noise can produce high reweighted SNR in LIGO's matched filter, well higher than the specified threshold of $5$ . Similar results are observed for GW151226 and GW170104.

This means, we cannot be sure whether GW signal or a low-amplitude sine wave or noise burst, produced high SNR in LIGO's matched filter. It is possible that there was \textbf{coincident} false detection at H1 and L1, due to a low-amplitude sine wave at one detector and noise burst in the other detector.  We do not know the probability of this false coincident detection, due to unknown external factors. LIGO's false alarm rate calculation is \textbf{not} applicable for this case, which pertains only to coincident false detection due to detector noise.

It should be noted that we \textbf{do not} claim that a sinewave or noise burst caused GW150914 or the other GW signals. The point is that when LIGO matched filter shows high SNR, it could be due to a sine wave or noise burst as well. We cannot be sure that a GW signal caused high SNR. 

On the other hand, normalized Cross Correlation Function(CCF) in time domain, described in the next subsection, of whitened and filtered $w_d(t) + a(t)$ with the template $h(t)$, CCF of $w_d(t) + b(t)$ with the template $h(t)$, and also CCF of $w_d(t) + a(t)$ with $w_d(t) + b(t)$, using short windows for the duration of the GW signal, does not produce peaky CCF~\footnote{By peaky CCF, we mean that the ratio, $R$, of the the absolute value of CCF for any lag greater than the decorrelation time of the template, to the absolute maximum value of CCF, should be less than a certain threshold. Decorrelation time of the template $\tau_{0}$ is defined as the time taken for the autocorrelation of the template to fall to $\frac{1}{e} = 0.36$ of the maximum value at zero lag~\citep{decorrelation_time} We will use the ratio $R_3$, which is the ratio of the absolute value of CCF at any lag greater than $\tau_0 *3$ to the absolute maximum value of CCF, and test whether $R_3 < \frac{1}{e}$. Lag greater than $\tau_0 *3$ is taken to allow for some cushion.} as in the lower panels in Fig.~\ref{fig:LIGO_matched_filter_sinewave}. Hence it is suggested that normalized CCF in time domain, using short windows be used. Besides, long 32 second windows used in LIGO's matched filter, include noise outside the duration of the GW signal, which can only reduce the sensitivity, as shown in the following subsections.

Hence H1 vs L1 cross-correlation test is necessary to avoid this wrong classification of low-amplitude noise burst/sine wave, as GW signals, detailed in Section~\ref{sec:test-3b}. 

Section~\ref{sec:false_alarm_rate} explains in detail why the false alarm rate calculation is not applicable for this example in the presence of non-stationary detector noise.

\subsection*{ Normalized CCF method}
\label{sec:normalized-ccf}

Let us consider a normalized Cross Correlation Function(CCF) described in Eq~\ref{eqn:3}, where both $s(t)$ and $h(t)$ are normalized over the time window [-T,T] during which GW signal was observed, such that CCF of each signal with itself gives a result of unity for zero lag. GW150914 was observed in the window of duration 0.2 seconds, GW151226 in a duration of 1 second and GW170104 in a duration of 0.12 seconds\citep{Ligo1,Ligo2, Ligo3}.\footnote{Normalized CCF using running windows method using, say 32 seconds of H1 and 0.2 seconds of L1, gives similar results as this method using short windows for both H1 and L1.} We will use reference systems as follows.

\begin{align}
\begin{split}
{CCF(t) =  \int_{-T}^{T} s(\tau) h(t-\tau) d\tau}
\end{split}
\label{eqn:3}
\end{align}

\textbf{Normalized CCF of H1/L1 with template is the same as "matched filtering of H1/L1 with template" using short windows in time domain and is an equally sensitive test.} It was shown in Section~\ref{sec:ligo_mf} that matched filter in fact does cross-correlation. The tests listed in this section are \textbf{new tests }with short windows.

In this manuscript, the term "Normalized CCF" is used to describe the cross-correlation of \textbf{any two signals}, such as H1/L1 with templates \textbf{or} H1 with L1, in time domain, using short windows over which GW event was observed, and the signals are normalized such that CCF of a signal with itself gives a value of unity at zero lag. The term "LIGO matched filter" is used to describe the matched filter implemented in LIGO python script\citep{LIGO Tutorials}, which does matched filtering of H1/L1 and the template, using 32 second windows, implemented in frequency domain. 

\subsection*{Long windows vs short windows for CCF}
\label{sec:short_windows}

\begin{figure}[h!]
\centering

\includegraphics[width=0.48\textwidth,height=0.14\textheight]{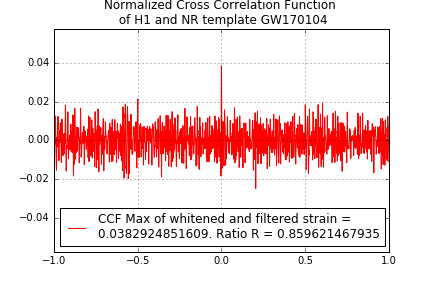}
\includegraphics[width=0.48\textwidth,height=0.14\textheight]{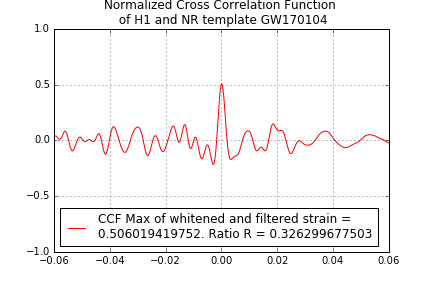}
\includegraphics[width=0.48\textwidth,height=0.14\textheight]{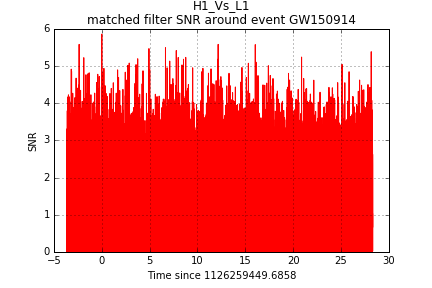}
\includegraphics[width=0.48\textwidth,height=0.14\textheight]{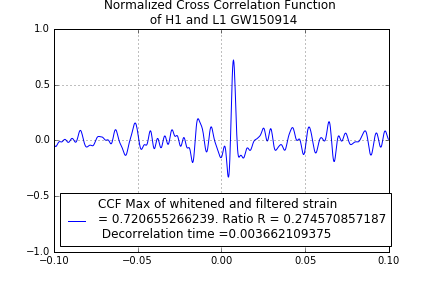}

\caption{Upper Panels: Normalized CCF for GW170104. Left Panel: done over long 20 second window. Right panel: done over GW event duration of  0.12 seconds. Lower Panels: GW150914: Left Panel: LIGO matched filter SNR H1 vs L1, done over 32 second windows. Right panel:Normalized CCF done over GW event duration of  0.2 seconds. If maximum value of CCF is negative, the plot is inverted. Ratio $R=R_3$.}
\label{fig:Normalized_CCF_plots_GW170104}
\end{figure}

GW150914, GW151226 and GW170104 have been observed over a window of duration 0.2, 1 and 0.12 seconds respectively\citep{Ligo1,Ligo2,Ligo3} and more than $95$ percent of the template energy is in this window and hence there is\textbf{ no point} in using windows longer than the duration of these signals.

The upper left panel in Fig.~\ref{fig:Normalized_CCF_plots_GW170104} shows normalized CCF done for GW170104, correlating H1 with the template, over long 20 second windows. We can see that the CCF is less peaky than the right panel and the decorrelation ratio(explained in Section~\ref{sec:test-3b}) is very high ($R > \frac{1}{e}$). On the other hand, the upper right panel in Fig.~\ref{fig:Normalized_CCF_plots_GW170104} correlates H1 with the template, over short 0.12 second window, over which the GW event was observed and CCF is sufficiently peaky and the decorrelation ratio $R < \frac{1}{e}$. 

Lower panels in Fig.~\ref{fig:Normalized_CCF_plots_GW170104} show similar results for GW150914. Lower left panel shows LIGO matched filter SNR, correlating H1 and L1, using long 32 second windows and do not show peaky CCF, while lower right panel performs normalized CCF over short 0.2 second windows, correlating H1 vs L1 and shows peaky CCF. Hence it is suggested that normalized CCF using short windows be used.

\subsection{Error in LIGO Whitening operation in the matched filter}
\label{sec:ligo_whitening}

\begin{figure}[h!]
\centering

\includegraphics[width=0.48\textwidth]{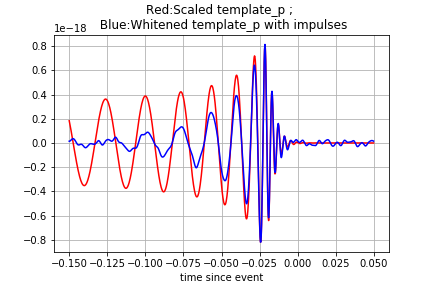}
\includegraphics[width=0.48\textwidth]{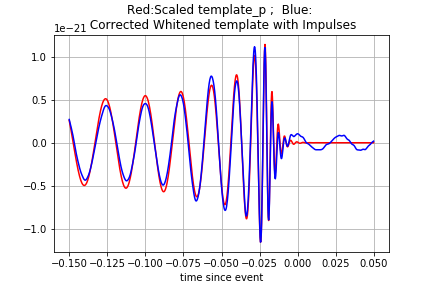}

\caption{GW150914:  Left Panel: LIGO Whitening of (ideal template + impulsive interference).    Right Panel: Modified Whitening only around impulsive interference of (ideal template + impulsive interference). }
\label{fig:ligo_whitening_1}
\end{figure}

LIGO detector noise has very high impulsive interference \href{https://losc.ligo.org/s/events/GW150914/P1500238/fig1.png}{here}, which is picked up from 60*n Hz power line harmonics, suspension violin modes and calibration test tones. The removal of this impulsive interference is called whitening.

LIGO software whitens the H1 and L1 strains $s(t)$ in the frequency domain, using $\frac{S(f)}{|S_{n}(f)|}$, across all frequencies, as detailed in Eq.1-4 in ~\citep{Ligo4}, where $S_{n}(f)$ is the power spectral density of detector noise and this introduces amplitude and phase distortion in the whitened strains.  This is illustrated in the left panel of Fig.~\ref{fig:ligo_whitening_1} using the example of an ideal template in the presence of impulsive interference in the detector noise \href{https://losc.ligo.org/s/events/GW150914/P1500238/fig1.png}{here}.\\

A modified whitening procedure is proposed here, where we whiten the strain, only around the 
vicinity of the impulsive interference. We can see in the right panel in Fig.~\ref{fig:ligo_whitening_1}  that this procedure works better than LIGO whitening procedure and lowers amplitude and phase distortion.\\

We can show mathematically how the whitening procedure used in LIGO python scripts given by $S_w(f)=\frac{S(f)}{|S_{n}(f)|}$ introduces amplitude and phase distortion in the whitened strains. This is illustrated in a simplified example, with $s(t) = h(t) + w_{em}(t)$ where $h(t)$ is the ideal template and $ w_{em}(t) = A \cos(2 \pi f_0 t)$ is the  60 Hz EM interference, where $f_0 = 60 + \delta$, where $\delta$ is chosen to be in between the FFT frequency bins, so that they will have an effect on the adjacent frequency bins. This is a reasonable example given that 60 Hz EM interference is likely to have a frequency spread around 60 Hz.

\begin{align}
\begin{split}
s(t) =  h(t) + w_{em}(t) = h(t) + A \cos(2 \pi f_0 t) \\
S(f) = H(f) +  W_{em}(f)  = (H_r(f) + i* H_i(f)) +  (W_r(f) + i * W_i(f)) \\
|S(f)| = \sqrt{(H_r(f)+W_r(f))^{2} + (H_i(f)+W_i(f))^{2} } \\
\end{split}
\label{eq1}
\end{align}

This can also be written as follows.

\begin{align}
\begin{split}
S(f) = H(f) +  W_{em}(f) = |H(f)| e^{i \theta_h(f)} + |W_{em}(f)| e^{i \theta_w(f)} \\
= [|H(f)| \cos{ \theta_h(f)} + |W_{em}(f)| \cos{ \theta_w(f)} ] + i [|H(f)| \sin{ \theta_h(f)} + |W_{em}(f)| \sin{ \theta_w(f)} ]   \\
S_w(f) =  \frac{S(f)}{|S_{n}(f)|} =   \frac{|S(f)|}{|S_{n}(f)|} e^{i \tan^{-1}{\frac{[|H(f)| \sin{ \theta_h(f)} + |W_{em}(f)| \sin{ \theta_w(f)} ] }{[|H(f)| \cos{ \theta_h(f)} + |W_{em}(f)| \cos{ \theta_w(f)} ]}} } \\
S_w(f) =  \frac{ \sqrt{(H_r(f)+W_r(f))^{2} + (H_i(f)+W_i(f))^{2} }}{|S_{n}(f)|} e^{i \tan^{-1}{\frac{[|H(f)| \sin{ \theta_h(f)} + |W_{em}(f)| \sin{ \theta_w(f)} ] }{[|H(f)| \cos{ \theta_h(f)} + |W_{em}(f)| \cos{ \theta_w(f)} ]}} }
\end{split}
\label{eq1}
\end{align}

Hence whitening using $S_w(f)=\frac{S(f)}{|S_{n}(f)|}$ causes amplitude distortion in the whitened strain, given that the magnitude of FFT of the ideal template has been artificially modified by the whitening procedure. 

In the absence of 60 Hz EM interference, $W_{em}(f) = 0$ and $S_w(f) =  \frac{|H(f)|}{|S_{n}(f)|} e^{i  \theta_h(f)}$ and we can see that the addition of 60 Hz EM interference causes phase distortion in the whitened strain. 

We know that the Fourier Transform of a continuous time signal $w_{em}(t)= A \cos(2 \pi f_0 t)$ is given by $ W_{em}(f) = \frac{A}{2} ( \delta(f-f_0) + \delta(f+f_0) )$. In discrete-time version, we have $W_{em}[k] = \displaystyle\sum_{n=0}^{N-1} w_{em}[n] e^{-i 2 \pi n \frac{k}{N}}$ and given that the chosen frequency $f_0$ could be in between the frequency bins, it will spread into adjacent bins and cause more amplitude and phase distortion.

\section{ Bogus Chirp Templates produce false peaks in the matched filter}
\label{sec:bogus_chirp}

\begin{figure}[h!]
\centering

\includegraphics[width=0.32\textwidth]{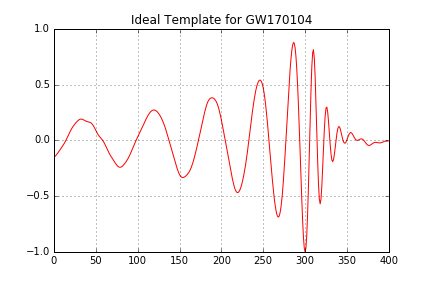}
\includegraphics[width=0.32\textwidth]{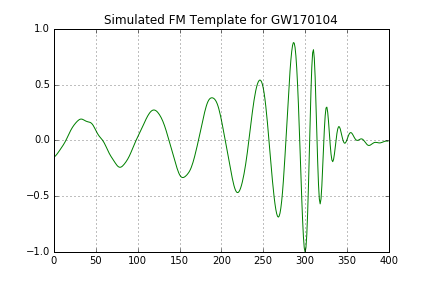}
\includegraphics[width=0.32\textwidth]{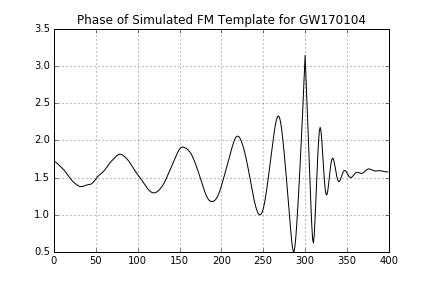}

\caption{GW170104:  Left:  Ideal template. Middle: Simulated FM Chirp Template. Right: Phase of Simulated FM Template }
\label{fig:bogus_chirp_0}
\end{figure}

Let us consider the ideal template $h(t)$ for GW170104, which is shown in the left panel of Fig.\ref{fig:bogus_chirp_0}.
We can simulate this ideal template by a frequency modulated(FM) waveform in the middle panel, where  $h_s(t) =  cos(m(t))$, where $m(t)$ is the phase of FM signal, which is shown in the right panel. This simulated FM signal also is a chirp signal with frequency increasing from the left to the right.

Now we can simulate bogus chirp templates $h_b(t)$, where $h_b(t) = cos( m_b(t))$, $ m_b(t)= m(t) + w_m(t)$ and $ w_m(t)$ is additive white gaussian noise(AWGN) added to the phase of the FM template.

We will inject these bogus chirp templates to the detector noise and show in the section below, that LIGO matched filter
misfires with high SNR, even for bogus chirp templates. We will show similar results using time domain CCF method.

It is possible that there was \textbf{coincident} false detection at H1 and L1, due to a bogus chirp template at one detector and a different bogus chirp template or a noise burst, in the other detector.  We do not know the probability of this false coincident detection, due to unknown external factors. LIGO's false alarm rate calculation is \textbf{not} applicable for this case, which pertains only coincident false detection due to detector noise.

Irrespective of the source of these bogus chirp templates, it will be argued that H1 vs L1 cross-correlation test is \textbf{necessary} to avoid this wrong classification of bogus templates as GW signals, detailed in Section~\ref{sec:test-3b}, given the need for high standards for classifying a GW signal.


\subsection*{False SNR peaks in LIGO matched filter with bogus templates for GW150914. }
\label{sec:test-1}

\begin{figure}[h!]
\centering

\includegraphics[width=0.48\textwidth]{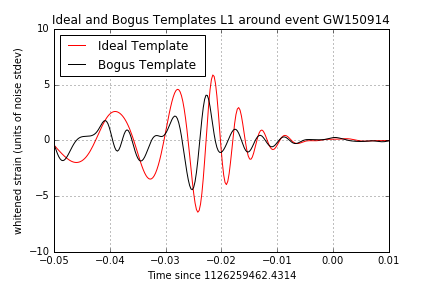}
\includegraphics[width=0.48\textwidth]{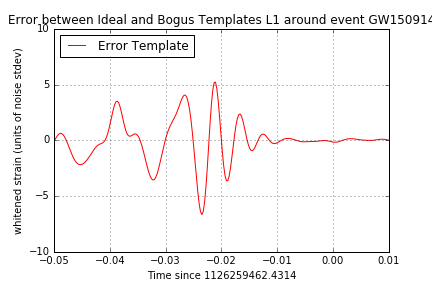}

\includegraphics[width=0.48\textwidth]{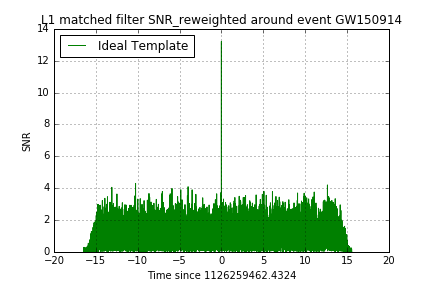}
\includegraphics[width=0.48\textwidth]{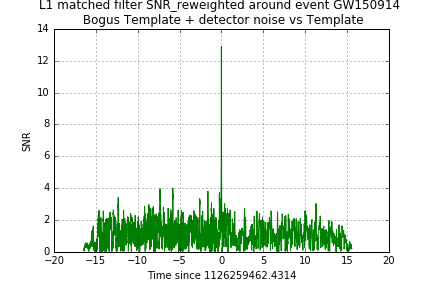}

\caption{GW150914:  Upper Left: Red: Ideal template. Black: Bogus Template. Upper Right: Error between Ideal and Bogus Templates. Lower Left: LIGO matched filter Reweighted SNR for L1 vs Ideal template. Lower Right: LIGO matched filter Reweighted  SNR for L1 vs Bogus template. }
\label{fig:ligo_mf_plots_0}
\end{figure}

Let us consider the case where  a noisy signal roughly resembling the template but generated by frequency modulation(FM)~\footnote{The template for GW150914 is represented by a FM signal $h(t) =  cos(m(t))$. Bogus template is given by $h_b(t) = cos( m_b(t))$ where $ m_b(t)= m(t) + w_m(t)$ and $ w_m(t)$ are white gaussian noise.}, is observed at L1. This bogus template also produces SNR peaks, when correlated with the template, as in the lower right panel of Fig.~\ref{fig:ligo_mf_plots_0}. This happens more than 50 percent of the time, in Monte-Carlo simulations with white gaussian noise.

The upper left panel in Fig.~\ref{fig:ligo_mf_plots_0} shows the ideal L1 template of GW150914 and the bogus L1 template. The upper right panel shows the error between the L1 bogus template and the ideal template. The lower left panel in Fig.~\ref{fig:ligo_mf_plots_0} shows LIGO matched filter reweighted SNR, when correlating L1 with the template of GW150914 and the lower right panel shows SNR peaks obtained by correlating template vs L1 bogus template added to the detector noise.  We can see that GW150914 shows SNR peaks \textbf{comparable} with SNR peaks corresponding to bogus template correlations. 

Given that bogus template can produce comparable correlations with the template, as the observed GW150914 signal, this means that we \textbf{cannot be sure} whether the SNR peaks in GW150914 were caused by GW signals or bogus template correlations. Hence H1 vs L1 cross-correlation test is necessary to avoid this wrong classification of bogus templates as GW signals, detailed in Section~\ref{sec:test-3b}.

Similar results are observed for GW151226 and GW170104 and are detailed in the section~\ref{sec:appendix-a}.


\section{Need for H1 vs L1 cross-correlation test: Ruling Out False detection of non-GW signals}
\label{sec:test-3b}

\begin{figure}[h!]
\centering
\includegraphics[width=0.32\textwidth]{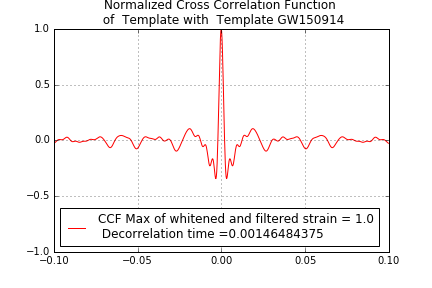}
\includegraphics[width=0.32\textwidth]{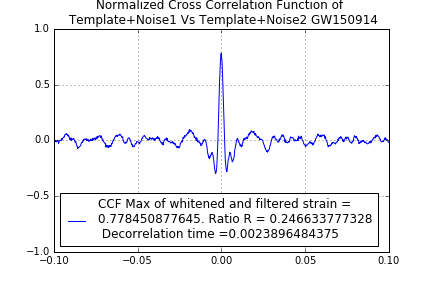}
\includegraphics[width=0.32\textwidth]{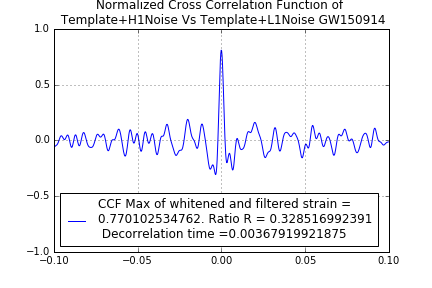}

\caption{ Reference Systems with Normalized CCF over 0.2 second windows for GW150914. Left:  Reference System A. Middle:  Reference System 1.  Right: Reference System 2. If maximum value of CCF is negative, the plot is inverted.}
\label{fig:Normalized_CCF_plots_00}
\end{figure}

\begin{figure}[h!]
\centering

\includegraphics[width=0.48\textwidth]{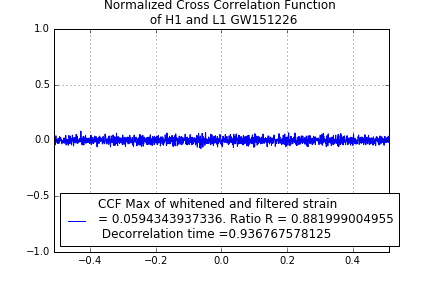}
\includegraphics[width=0.48\textwidth]{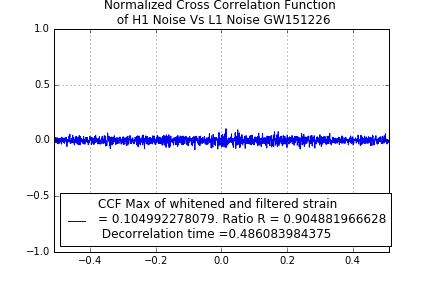}

\includegraphics[width=0.48\textwidth]{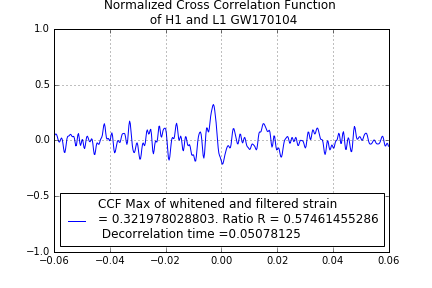}
\includegraphics[width=0.48\textwidth]{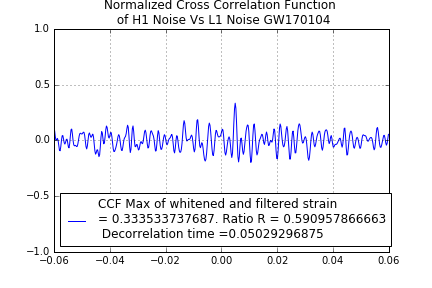}

\caption{Normalized CCF Plots done over GW event duration. Top Row: GW151226.   Bottom Row: GW170104. Left panels: H1 vs L1. Right panels: H1 noise vs L1 noise. H1 and L1 detector noise obtained 1 second after the end of GW151226 and 10 seconds after the end of GW170104.}
\label{fig:Normalized_CCF_plots_2}
\end{figure}

We do not want to accept above false coincident detection in section~\ref{sec:ligo_mf}, section~\ref{sec:bogus_chirp} as a valid GW signal, because they were false coincidences. We can rule out these cases of false coincident detection by correlating H1 with L1, which gives poor CCF peaks for GW151226 and GW170104, as in the left panel in Fig.~\ref{fig:Normalized_CCF_plots_2}, which are indistinguishable from detector noise correlations in the right panels. 

It is possible that there was \textbf{coincident} false detection at H1 and L1, due to any combination of sine wave/noise burst/bogus chirp templates/detector noise, in both detectors.  We do not know the probability of this false coincident detection, due to unknown external factors. LIGO's false alarm rate calculation is \textbf{not} applicable for this case, which pertains to only coincident false detection due to detector noise.

Given that the same GW signal is expected to be received in both sites, these signals should give a high CCF when correlated with each other. This is a \textbf{crucial} test which must be performed.

We\textbf{ can }cross-correlate two noisy signals and expect a peaky CCF, if the signals are correlated. In fact, wireless communication with sensors routinely use cross-correlation of two noisy signals~\citep{Garnier}. We wish to correlate two noisy signals $s(t)=h_H(t)+w_H(t)$[H1] and  $h^{'}(t)= h_L(t)+w_L(t)$[L1] where $w_H(t)$ and $w_L(t)$ represent the detector noise, and $h_H(t)$ and $h_L(t)$ represent the templates of H1 and L1. This is equivalent to correlating $s(t)=h^{'}(t)+(w_H(t)-w_L(t))+(h_H(t)-h_L(t))$ and  $h^{'}(t)$ and hence correlating $s(t)= h^{'}(t) + w(t)$ and  $h^{'}(t)$, where $w(t)=(w_H(t)-w_L(t))+(h_H(t)-h_L(t))$. $h^{'}(t)$ is the noisy template. The theory of matched filter or cross-correlation imposes no constraints on the characteristics of the template. 

The same template $h(t)$ is used in the right panels CCF plots of  Fig.~\ref{fig:Normalized_CCF_plots_00}, because the templates received at H1 and L1 detectors are nearly identical, which differ only by a time shift and a scale factor, and this does not matter for the ratio $R$, in the the normalized CCF plots. We observe that $R < \frac{1}{e}$. Given that GW signals are assumed to be the sum of an ideal template and detector noise, this comparison is reasonable.

\textbf{Reference System A:} An ideal template for GW151226 or GW170104 $h(t)$ is correlated with itself, after normalization, to give a peaky CCF. The maximum value of CCF = 1 at zero lag, as in left panels in Fig.~\ref{fig:Normalized_CCF_plots_00}.

\textbf{Reference System 1:}  Fig.~\ref{fig:Normalized_CCF_plots_00} shows this system in the second panel, which correlates two noisy signals, $s(t)$ and $h^{'}(t)$, where $s(t)$ is the sum of the template for GW150914 and AWGN, and $h^{'}(t)$ is the sum of the same template and an independent AWGN. Then it computes the normalized ${{CCF(t) = \int_{-T}^{T} s(\tau) h^{'}(t-\tau) d\tau}}$. We can see that the CCF shows strong peaky behaviour. Average decorrelation time for this system is $\tau_{1} = 0.0024$ seconds.

\textbf{Reference System 2:}  Fig.~\ref{fig:Normalized_CCF_plots_00} shows a reference system 2 in the third panel, which correlates two noisy signals, $s(t)$ and $h^{'}(t)$, where $s(t)$ is the sum of the template for GW150914 and H1 detector noise outside the GW event, and $h^{'}(t)$ is the sum of the same template and and L1 detector noise outside the GW event, and computes the normalized CCF. We can see that the CCF shows strong peaky behaviour. Average decorrelation time for this system is $\tau_{2} = 0.0037$ seconds. 

Reference systems 1 and 2 are shown only for the purpose of demonstrating the fact that we \textbf{can} cross-correlate two noisy signals and expect a peaky CCF. Given the need for high standards required in classifying GW signals, we will use only the decorrelation time $\tau_0 $ of the reference system A, which correlates the template of each GW signal with itself, when we compare the decorrelation times and the ratio R, of the three GW signals.

Fig.~\ref{fig:Normalized_CCF_plots_2} plots the normalized CCF for GW151226(top row) and GW170104 (bottom row) by correlating H1 with L1 in the left column and correlating  H1 detector noise with L1 detector noise in the right column. We can see that GW151226 and GW170104 show very poor CCF peaks ($R_3 > \frac{1}{e}$) when correlating H1 with L1, and we can see that CCF peaks are \textbf{indistinguishable} from CCF peaks corresponding to detector noise in the right column. Hence, we cannot be sure whether detector noise or GW signal caused the CCF peaks.

We can reproduce these results for \textbf{longer H1/L1 strains} of GW151226 and GW170104 as well. \href{https://www.ocf.berkeley.edu/~araman/files/ligo_mf/GW150914/figures/Fig-a3.png}{Figure} shows CCF done over 14 second duration, when correlating H1 with L1 for GW151226(top panel) and GW170104(bottom panel).  We can see that GW151226 and GW170104 show CCF peaks (left panel) \textbf{indistinguishable} from CCF peaks corresponding to detector noise correlations (right panel).

Hence \textbf{GW170104} and \textbf{GW151226} should be further studied, based on this test alone[Reason 5].

\section{ H1 vs L1 cross-correlation test fails for GW170814 and GW170608}
\label{sec:test-3c}

\begin{figure}[h!]
\centering

\includegraphics[width=0.48\textwidth]{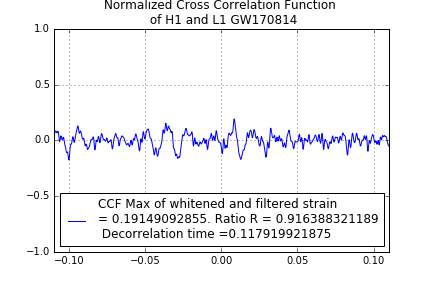}
\includegraphics[width=0.48\textwidth]{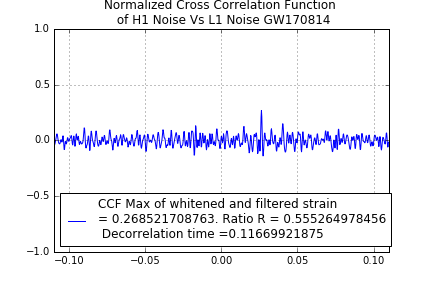}

\includegraphics[width=0.48\textwidth]{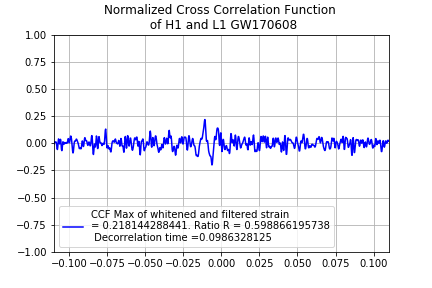}
\includegraphics[width=0.48\textwidth]{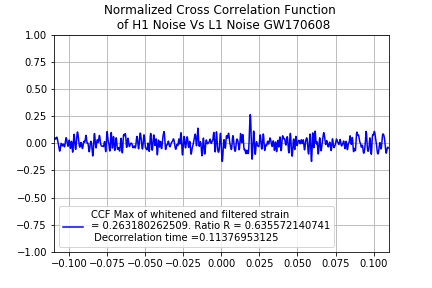}

\caption{Normalized CCF Plots done over GW event duration of 0.2 seconds. Top Row: GW170814.   Bottom Row: GW170608. Left panels: H1 vs L1. Right panels: H1 noise vs L1 noise. H1 and L1 detector noise obtained 10 seconds after the end of GW events.}
\label{fig:Normalized_CCF_plots_3}
\end{figure}

GW170814 has been observed at three detectors H1, L1 and V1. For the purpose of this section, we will do the correlation test for H1 vs L1 only and this is enough to show that the correlation test fails.

Fig.~\ref{fig:Normalized_CCF_plots_3} plots the normalized CCF for GW170814(top row) and GW170608 (bottom row) by correlating H1 with L1 in the left column and correlating  H1 detector noise with L1 detector noise in the right column. We can see that GW170814 and GW170608 show very poor CCF peaks ($R_3 > \frac{1}{e}$) when correlating H1 with L1, and we can see that CCF peaks are \textbf{indistinguishable} from CCF peaks corresponding to detector noise in the right column. Hence, we cannot be sure whether detector noise or GW signal caused the CCF peaks.

Hence \textbf{GW170814} and \textbf{GW170608} should be further studied, based on this test alone[Reason 6].

\section{Electro-Magnetic Interference in the GW channel as a candidate for GW signals}
\label{sec:GW170814_GW170608}

The gravitational wave(GW) channel in the LIGO detectors, picks up 60*n Hz electro-magnetic(EM) interference(\href{https://losc.ligo.org/s/events/GW150914/P1500238/fig1.png}{figure}) and also wideband EM interference. EM interference may enter the GW channel through several paths, by air or wires. Each such path in the GW channel may have a different coupling (frequency response).

It is quite possible that lightning, which is an EM interference signal, can be picked up at both detectors H1 and L1 and may be mistaken for a GW signal. Lightning has \href{https://theinspireproject.org/default.asp?contentID=4}{low frequency} components, in the  0 to 2048 Hz range. EM signals also travel at the speed of light, similar to the GW signal, and arrive within the 10 ms window at the two detectors.

Given that the strongest signal GW150914 is buried by a factor of $\frac{1}{400}$ in the raw detector noise in Fig.\ref{fig:ligo_whitening_0} and does not show any signal spikes in the time domain, the raw magnetometer signal in time domain, also may not show any signal spikes during the GW events.

We can rule out EM interference in the GW channel, by doing cross-correlation of the magnetometer signal with the template, during GW event duration. If the EM signal in the magnetometer channel is subjected to a transfer function different from that of the GW channel, it is entirely possible that the magnetometer channel may not produce cross correlation peaks when correlated with the template and hence this EM signal may be mistaken for GW signals.

This possibility is explored in detail \href{https://www.ocf.berkeley.edu/~araman/files/ligo_tests/ligo_EM_v2.pdf}{here} and shown that EM interference is a good candidate for all the 6 GW signals observed so far.

\subsection{The case of GW170817}
\label{sec:GW170814_GW170817}

The Gamma Ray Burst(GRB) which was observed with GW170817 was picked up by the telescopes. We know that astrophysical objects such as stars, emit EM signals in a broad frequency range, down to KHz. Our Sun is also known to emit EM signals as low as 30 KHz (NASA \href{https://swaves.gsfc.nasa.gov/Vocabulary%20SWAVES.html}{document}).  We know that \href{https://www.ocf.berkeley.edu/~araman/files/ligo_tests/GW170817/GW170817_cln_not_superimposed_1.png}{GW170817} was observed as a weak signal, whose amplitude does not rise above the detector noise during the GW event. If the astrophysical object which emitted the Gamma Ray Burst, also emitted EM signal in the 0 to 2048 Hz frequency range, magnetometers may not show any signal spikes in time domain during the GW event and hence low frequency EM signals from the astrophysical object which emitted the GRB, may be a good candidate for GW170817.

\section{Concluding remarks}\label{conclusion}
It is possible that there was \textbf{coincident} false detection at H1 and L1, due to any combination of external sources of sine wave/noise burst/bogus chirp templates, in both detectors. We do not know the probability of this false coincident detection, due to unknown external factors. LIGO's false alarm rate calculation is \textbf{not} applicable for this case, which pertains to only coincident false detection due to detector noise. It is also possible that external EM signals like lightning may be picked up in the GW channel and mistaken as GW.

Reiterating the point made earlier, because we are more often likely to observe weak signals which look like noise and whose amplitude does not rise during assumed GW event, it is of paramount importance that we should not classify noise or EM interference as GW events. We need \textbf{high} standards to classify an observed time series as a GW signal. 

\acknowledgments
I am grateful to Andrew D. Jackson, Dr. A. Paulraj, Helmut Bolcskei and John M Cioffi for encouragement, suggestions and discussions. I would like to thank Sebastian Domenico von Hausegger and Arunava Chaudhuri for review of our Python scripts and helpful suggestions. I would like to thank M.A. Srinivas, Hao Liu, James Creswell, Bhavna Antony, Anant Sahai and Kannan Ramachandran for discussions and helpful suggestions. I would like to thank LIGO Open Science Center for making the data and Python scripts available online. I would like to thank LIGO scientists who answered  many questions in detail.

\clearpage

\section{Appendix A}\label{sec:appendix-a}

\subsection*{False SNR peaks in LIGO matched filter with bogus templates for GW151226. }
\label{sec:test-3}

\begin{figure}[h!]
\centering

\includegraphics[width=0.48\textwidth]{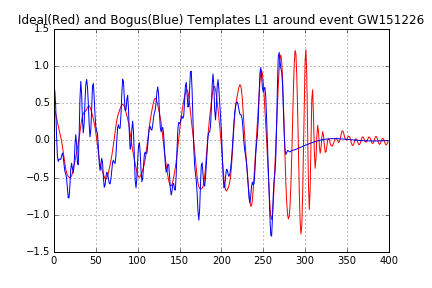}
\includegraphics[width=0.48\textwidth]{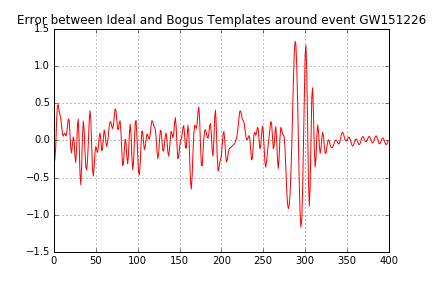}

\includegraphics[width=0.48\textwidth]{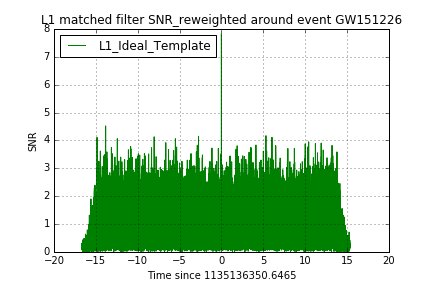}
\includegraphics[width=0.48\textwidth]{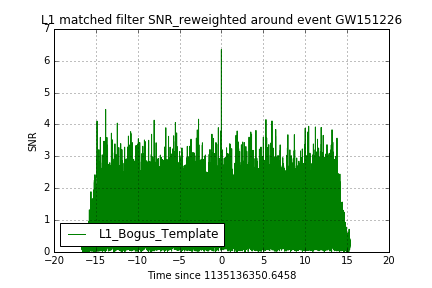}

\caption{GW151226:  Upper Left: Red: Ideal template. Black: Bogus Template. Upper Right: Error between Ideal and Bogus Templates. Lower Left: LIGO matched filter Reweighted SNR for L1 vs Ideal template. Lower Right: LIGO matched filter Reweighted SNR for L1 vs Bogus template. }
\label{fig:ligo_mf_plots_2}
\end{figure}

Let us consider the case where  a noisy signal roughly resembling the template but generated by frequency modulation(FM), is observed at L1. This bogus template $h_b(t)$ also produces SNR peaks, when correlated with the template, as in the lower right panel of Fig.~\ref{fig:ligo_mf_plots_2}. 

The upper left panel in Fig.~\ref{fig:ligo_mf_plots_2} shows the ideal template of GW151226 and the bogus L1 template. We can see that the bogus template differs \textbf{significantly} from the ideal template, in fact the last 5 cycles in the crucial final chirp portion is \textbf{completely absent}. The upper right panel shows the error between the L1 bogus template and the ideal template. The lower left panel in Fig.~\ref{fig:ligo_mf_plots_2} shows LIGO matched filter reweighted SNR, when correlating L1 with the template of GW151226 and the lower right panel shows SNR peaks obtained by correlating template vs L1 bogus template added to the detector noise.  We can see that the reweighted SNR due to the bogus template is greater than the threshold of 5 and GW151226 shows SNR peaks \textbf{comparable} with SNR peaks corresponding to bogus template correlations. 

We can also show high SNR peaks with a bogus H1 template, which is different from a bogus L1 template, as in Fig.~\ref{fig:ligo_mf_plots_12}. This may result in false coincidence. In Section~\ref{sec:test-6}, it is shown that detector noise itself may produce false cross-correlation peaks in L1, while a bogus template may produce false peaks in H1. 

Given that bogus template can produce comparable correlations with the template, as the observed GW151226 signal, this means that we \textbf{cannot be sure} whether the SNR peaks in GW151226 were caused by GW signals or bogus template correlations.
Hence H1 vs L1 cross-correlation test is necessary to avoid this wrong classification of bogus templates as GW signals, detailed in Section~\ref{sec:test-3b}. 

Using normalized CCF method in time domain, we can also show high CCF peaks with a bogus L1 template, which is different from a bogus L1 template, as in Fig.~\ref{fig:ligo_mf_plots_4} for GW151226 for GW170104.

Hence  \textbf{GW151226} should be further studied, based on this test alone[Reason 3c].

\subsection*{False SNR peaks in LIGO matched filter with bogus templates for GW170104. }
\label{sec:test-2}

\begin{figure}[h!]
\centering

\includegraphics[width=0.48\textwidth]{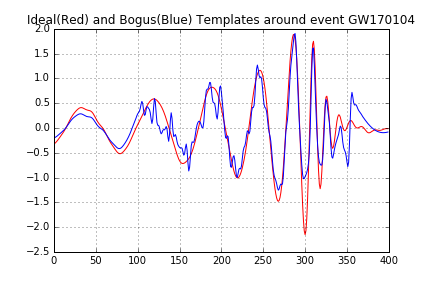}
\includegraphics[width=0.48\textwidth]{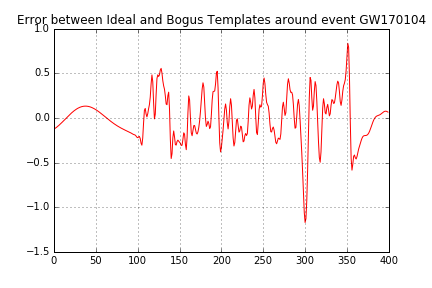}

\includegraphics[width=0.48\textwidth]{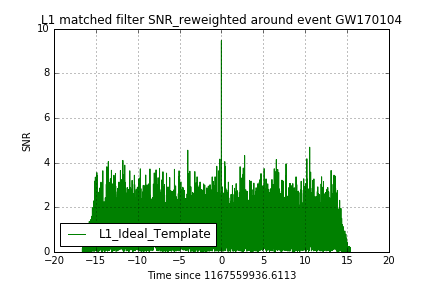}
\includegraphics[width=0.48\textwidth]{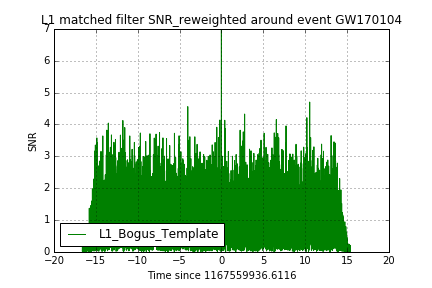}

\caption{GW170104:  Upper Left: Red: Ideal template. Black: Bogus Template. Upper Right: Error between Ideal and Bogus Templates. Lower Left: LIGO matched filter Reweighted SNR for L1 vs Ideal template. Lower Right: LIGO matched filter Reweighted  SNR for L1 vs Bogus template. }
\label{fig:ligo_mf_plots_1}
\end{figure}

Let us consider the case where  a noisy signal roughly resembling the template but generated by frequency modulation(FM), is observed at L1. This bogus template $h_b(t)$ also produces SNR peaks, when correlated with the template, as in the lower right panel of Fig.~\ref{fig:ligo_mf_plots_1}. 

The upper left panel in Fig.~\ref{fig:ligo_mf_plots_1} shows the ideal template of GW170104 and the bogus L1 template. The upper right panel shows the error between the L1 bogus template and the ideal template. The lower left panel in Fig.~\ref{fig:ligo_mf_plots_1} shows LIGO matched filter reweighted SNR, when correlating L1 with the template of GW170104 and the lower right panel shows SNR peaks obtained by correlating  template vs L1 bogus template added to the detector noise.  We can see that the reweighted SNR due to the bogus template is greater than the threshold of 5 and GW170104 shows SNR peaks \textbf{comparable} with SNR peaks corresponding to bogus template correlations. 

We can also show high SNR peaks with a bogus H1 template, which is different from a bogus L1 template, as in Fig.~\ref{fig:ligo_mf_plots_11}. This may result in false coincidence.

Given that bogus template can produce comparable correlations with the template, as the observed GW170104 signal, this means that we \textbf{cannot be sure} whether the SNR peaks in GW170104 were caused by GW signals or bogus template correlations.
Hence H1 vs L1 cross-correlation test is necessary to avoid this wrong classification of bogus templates as GW signals, detailed in Section~\ref{sec:test-3b}. 

Using normalized CCF method in time domain, we can also show high CCF peaks with a bogus L1 template, which is different from a bogus L1 template, as in Fig.~\ref{fig:ligo_mf_plots_3} for GW170104.

Hence  \textbf{GW170104} should be further studied, based on this test alone[Reason 3b].

\section*{ False peaks in time domain cross-correlation method due to Detector Noise and Bogus Chirp Templates}
\label{sec:false-peaks-0}

\subsection*{False coincidence with L1 detector noise and bogus template at H1, in 4096 second block of data. }
\label{sec:test-6}

\begin{figure}[h!]
\centering

\includegraphics[width=0.48\textwidth]{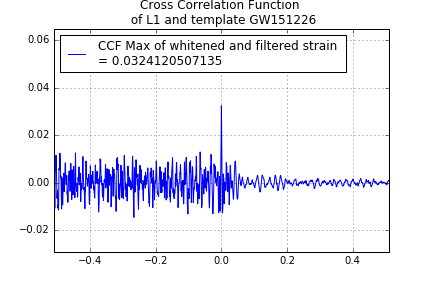}
\includegraphics[width=0.48\textwidth]{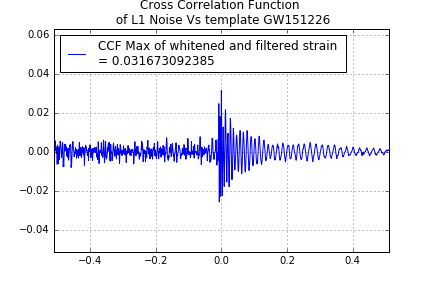}

\includegraphics[width=0.32\textwidth]{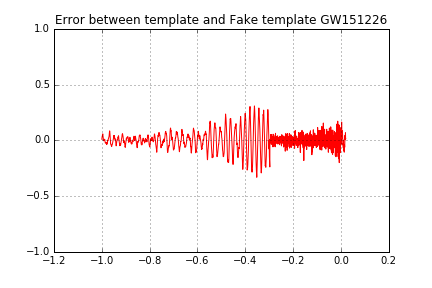}
\includegraphics[width=0.32\textwidth]{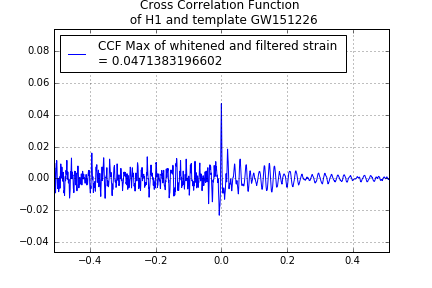}
\includegraphics[width=0.32\textwidth]{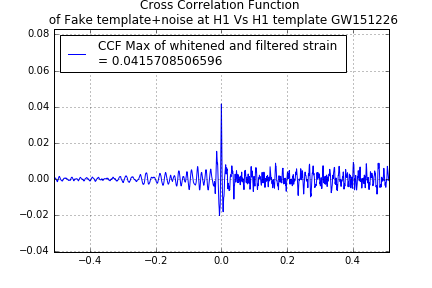}

\caption{GW151226: CCF Plots done over GW event duration of 1 second, using running windows for the right panels. Upper Left: L1 vs template. Upper Right: L1 Noise vs template. Lower Middle: H1 vs template. Lower Right: (Fake Template+H1 noise) vs template. Lower left: Error between actual and fake template.  H1/L1 detector noise obtained from a 4096 second block of data and excluding GW event portions and edges. }
\label{fig:CCF_plots_0}
\end{figure}

The upper left panel in Fig.~\ref{fig:CCF_plots_0} shows Cross Correlation Function (CCF) done over 1 second duration, when correlating L1 with the template of GW151226 and the upper right panel shows CCF peaks obtained by correlating  L1 detector noise with the template, using running windows, over 4096 second block of data, excluding the GW event portions and edges.  We can see that GW151226 shows CCF peaks \textbf{comparable} with CCF peaks corresponding to detector noise correlations, and \textbf{do not} rise well above detector noise correlations.

Given that detector noise can produce comparable correlations with the template, as the observed GW151226 signal, let us consider the case where a) detector noise at L1 produces CCF peaks, when correlated with the template, in 4096 second block of data, as in the upper right panel of Fig.~\ref{fig:CCF_plots_0}, and b) a noisy signal roughly resembling the template but generated by frequency modulation(FM) and amplitude modulation(AM)~\footnote{The template for GW151226 is represented by a FM+AM signal $h(t) = A(t) * cos(2 \pi f_0 t + m(t))$ where $f_0=56$ Hz. Bogus template is given by $h_b(t) = A_b(t) * cos(2 \pi f_0 t + m_b(t))$ where $A_b(t) = A(t)+ w_a(t), m_b(t)= m(t) + w_m(t)$ and $w_a(t), w_m(t)$ are white gaussian noise.} in the last 0.3 seconds and zero in the first 0.7 seconds, is observed at H1, which also produces CCF peaks, when correlated with the template, as in the lower right panel of Fig.~\ref{fig:CCF_plots_0}. 

It is possible that there was \textbf{coincident} false detection at H1 and L1, due to a bogus chirp template at one detector and detector noise, in the other detector.  We do not know the probability of this false coincident detection, due to unknown external factors. LIGO's false alarm rate calculation is \textbf{not} applicable for this case, which pertains only coincident false detection due to detector noise.

This means that we \textbf{cannot be sure} whether the CCF peaks in GW151226 were caused by GW signals or detector noise correlations and bogus templates. Similar results hold for the case of GW170104. Hence H1 vs L1 cross-correlation test is necessary to avoid this wrong classification of bogus templates as GW signals, detailed in Section~\ref{sec:test-3b}. Hence \textbf{GW170104} and \textbf{GW151226} should be further studied, based on this test alone[Reason 4c].

\subsection*{False peaks in time domain CCF with bogus templates for L1 for GW151226. }
\label{sec:test-5}

\begin{figure}[h!]
\centering

\includegraphics[width=0.48\textwidth]{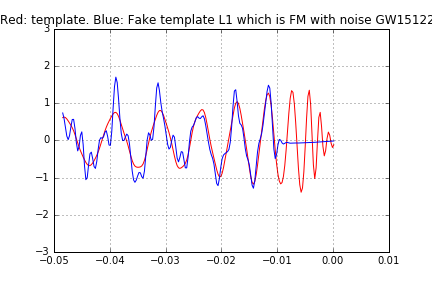}
\includegraphics[width=0.48\textwidth]{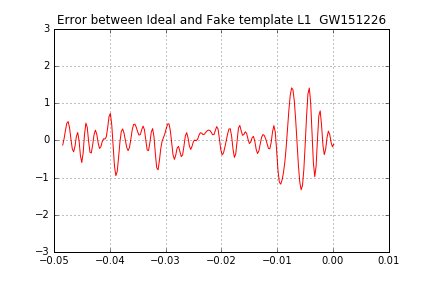}

\includegraphics[width=0.48\textwidth]{figure/GW151226_L1_template_CCF_2.png}
\includegraphics[width=0.48\textwidth]{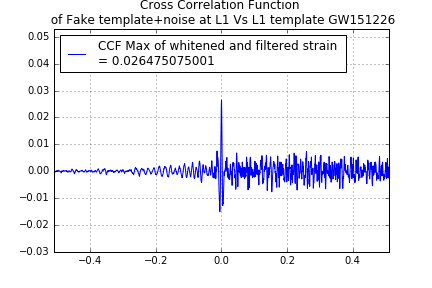}

\caption{GW151226:  Upper Left: Red: Ideal template. Black: Bogus Template. Upper Right: Error between Ideal and Bogus Templates. Lower Left: Normalized CCF in time domain for L1 vs Ideal template. Lower Right: Normalized CCF in time domain for L1 vs Bogus template. }
\label{fig:ligo_mf_plots_4}
\end{figure}

Let us consider the case where  a noisy signal roughly resembling the template but generated by frequency modulation(FM), is observed at L1. This bogus template $h_b(t)$ also produces CCF peaks, when correlated with the template, as in the lower right panel of Fig.~\ref{fig:ligo_mf_plots_4}. 

The upper left panel in Fig.~\ref{fig:ligo_mf_plots_4} shows the ideal template of GW151226 and the bogus L1 template. The upper right panel shows the error between the L1 bogus template and the ideal template. The lower left panel in Fig.~\ref{fig:ligo_mf_plots_4} shows Cross Correlation Function (CCF) done over 1 second duration, when correlating L1 with the template of GW151226 and the lower right panel shows CCF peaks obtained by correlating template vs L1 bogus template added to the detector noise.  We can see that GW151226 shows CCF peaks \textbf{comparable} with CCF peaks corresponding to bogus template correlations. 

We can also show high SNR peaks with a bogus H1 template, which is different from a bogus L1 template, as in Fig.~\ref{fig:ligo_mf_plots_14}. This may result in false coincidence.

Given that bogus template can produce comparable correlations with the template, as the observed GW151226 signal, this means that we \textbf{cannot be sure} whether the CCF peaks in GW151226 were caused by GW signals or bogus template correlations. Hence H1 vs L1 cross-correlation test is necessary to avoid this wrong classification of bogus templates as GW signals, detailed in Section~\ref{sec:test-3b}. Hence  \textbf{GW151226} should be further studied, based on this test alone[Reason 4a].

\subsection*{False peaks in time domain CCF with bogus templates for GW170104. }
\label{sec:test-4}

\begin{figure}[h!]
\centering

\includegraphics[width=0.48\textwidth]{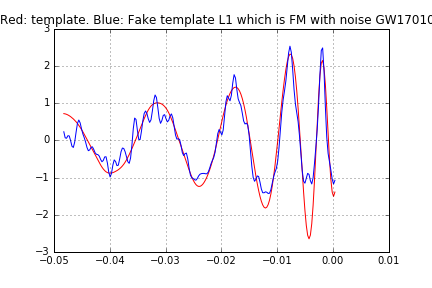}
\includegraphics[width=0.48\textwidth]{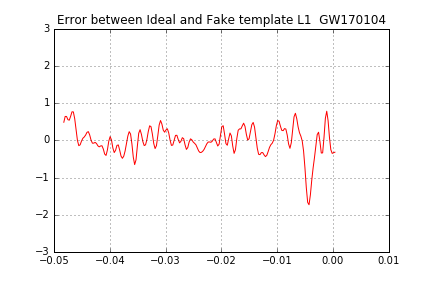}

\includegraphics[width=0.48\textwidth]{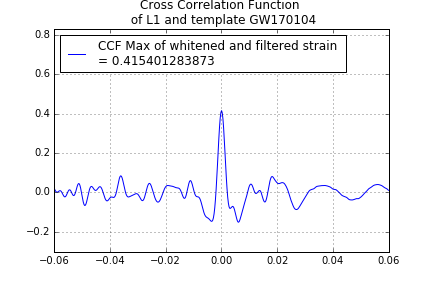}
\includegraphics[width=0.48\textwidth]{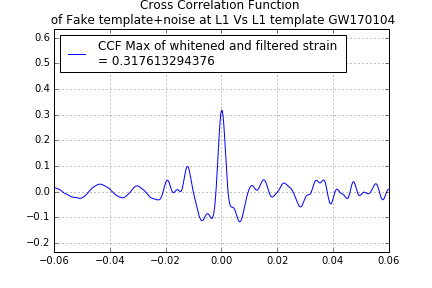}

\caption{GW170104:  Upper Left: Red: Ideal template. Black: Bogus Template. Upper Right: Error between Ideal and Bogus Templates. Lower Left: Normalized CCF in time domain for L1 vs Ideal template. Lower Right: Normalized CCF in time domain for L1 vs Bogus template. }
\label{fig:ligo_mf_plots_3}
\end{figure}

Let us consider the case where  a noisy signal roughly resembling the template but generated by frequency modulation(FM), is observed at L1. This bogus template $h_b(t)$ also produces CCF peaks, when correlated with the template, as in the lower right panel of Fig.~\ref{fig:ligo_mf_plots_3}. 

The upper left panel in Fig.~\ref{fig:ligo_mf_plots_3} shows the ideal template of GW170104 and the bogus L1 template. The upper right panel shows the error between the L1 bogus template and the ideal template. The lower left panel in Fig.~\ref{fig:ligo_mf_plots_3} shows Cross Correlation Function (CCF) done over 0.1 second duration, when correlating L1 with the template of GW170104 and the lower right panel shows CCF peaks obtained by correlating template vs  L1 bogus template added to the detector noise.  We can see that GW170104 shows CCF peaks \textbf{comparable} with CCF peaks corresponding to bogus template correlations. 

We can also show high SNR peaks with a bogus H1 template, which is different from a bogus L1 template, as in Fig.~\ref{fig:ligo_mf_plots_13}. This may result in false coincidence.

Given that bogus template can produce comparable correlations with the template, as the observed GW170104 signal, this means that we \textbf{cannot be sure} whether the CCF peaks in GW170104 were caused by GW signals or bogus template correlations. Hence H1 vs L1 cross-correlation test is necessary to avoid this wrong classification of bogus templates as GW signals, detailed in Section~\ref{sec:test-3b}. Hence  \textbf{GW170104} should be further studied, based on this test alone[Reason 4b].

\section{Appendix B}\label{appendix-b}

\subsection*{False peaks in LIGO matched filter  and time domain CCF due to bogus templates at H1, for GW152226, GW170104.}
\label{sec:false_peaks_H1}

\begin{figure}[h!]
\centering

\includegraphics[width=0.48\textwidth]{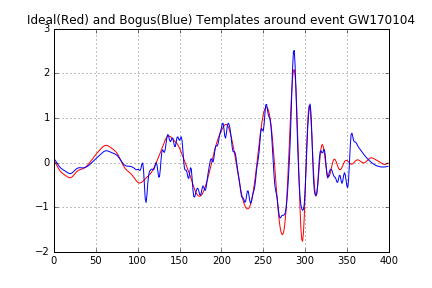}
\includegraphics[width=0.48\textwidth]{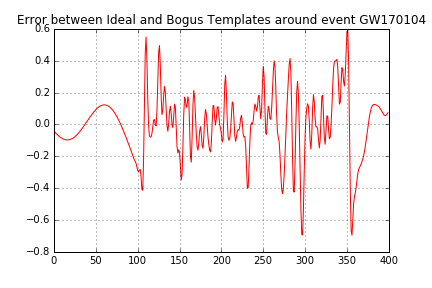}

\includegraphics[width=0.48\textwidth]{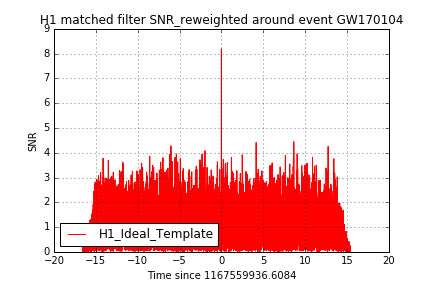}
\includegraphics[width=0.48\textwidth]{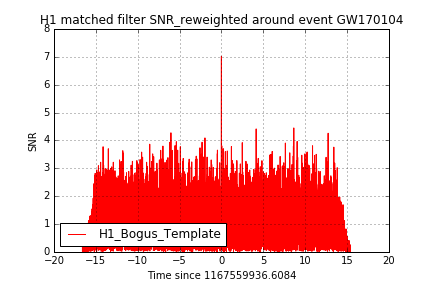}

\caption{GW170104:  Upper Left: Red: Ideal template. Black: Bogus Template. Upper Right: Error between Ideal and Bogus Templates. Lower Left: LIGO matched filter SNR for H1 vs Ideal template. Lower Right: LIGO matched filter SNR for H1 vs Bogus template. }
\label{fig:ligo_mf_plots_11}
\end{figure}

\begin{figure}[h!]
\centering

\includegraphics[width=0.48\textwidth]{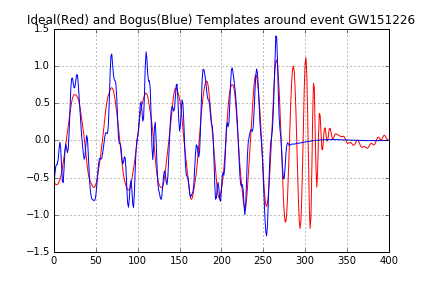}
\includegraphics[width=0.48\textwidth]{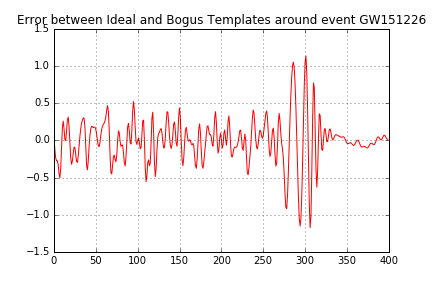}

\includegraphics[width=0.48\textwidth]{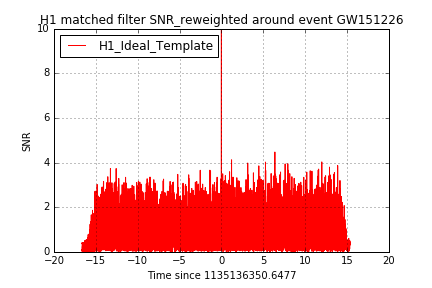}
\includegraphics[width=0.48\textwidth]{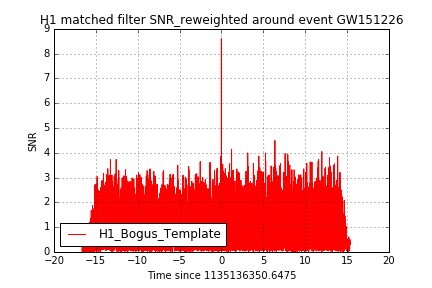}

\caption{GW151226:  Upper Left: Red: Ideal template. Black: Bogus Template. Upper Right: Error between Ideal and Bogus Templates. Lower Left: LIGO matched filter SNR for H1 vs Ideal template. Lower Right: LIGO matched filter SNR for H1 vs Bogus template. }
\label{fig:ligo_mf_plots_12}
\end{figure}

\begin{figure}[h!]
\centering

\includegraphics[width=0.48\textwidth]{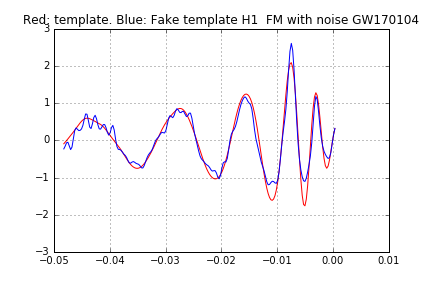}
\includegraphics[width=0.48\textwidth]{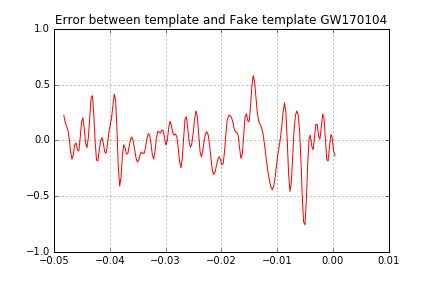}

\includegraphics[width=0.48\textwidth]{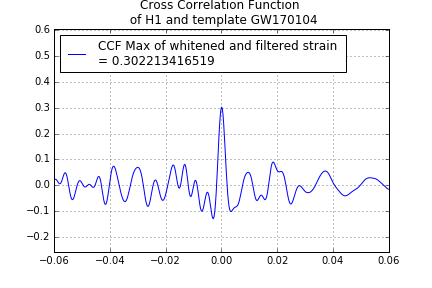}
\includegraphics[width=0.48\textwidth]{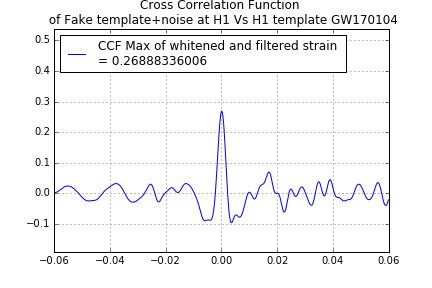}

\caption{GW170104:  Upper Left: Red: Ideal template. Black: Bogus Template. Upper Right: Error between Ideal and Bogus Templates. Lower Left: Normalized CCF in time domain for H1 vs Ideal template. Lower Right: Normalized CCF in time domain for H1 vs Bogus template. }
\label{fig:ligo_mf_plots_13}
\end{figure}

\begin{figure}[h!]
\centering

\includegraphics[width=0.48\textwidth]{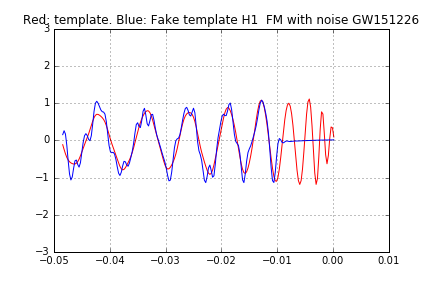}
\includegraphics[width=0.48\textwidth]{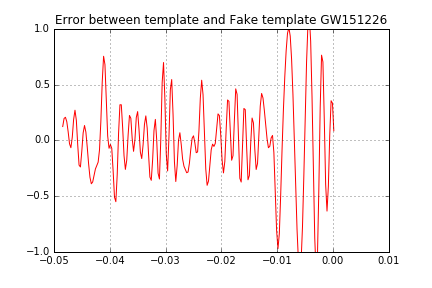}

\includegraphics[width=0.48\textwidth]{figure/GW151226_H1_template_CCF_2.png}
\includegraphics[width=0.48\textwidth]{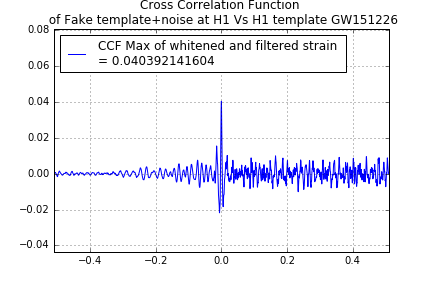}

\caption{GW151226:  Upper Left: Red: Ideal template. Black: Bogus Template. Upper Right: Error between Ideal and Bogus Templates. Lower Left: Normalized CCF in time domain for H1 vs Ideal template. Lower Right: Normalized CCF in time domain for H1 vs Bogus template. }
\label{fig:ligo_mf_plots_14}
\end{figure}

\clearpage

\subsection*{GW waveforms }
\label{sec:test-15}

\begin{figure}[h!]
\centering
\includegraphics[width=0.32\textwidth]{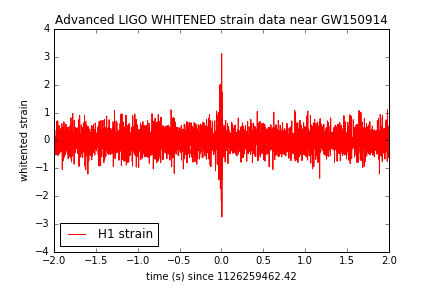}
\includegraphics[width=0.32\textwidth]{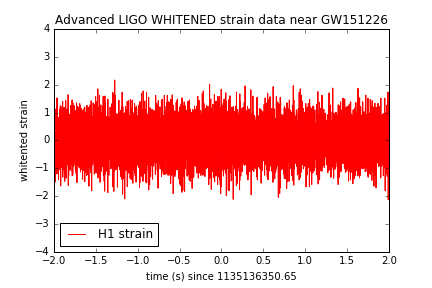}
\includegraphics[width=0.32\textwidth]{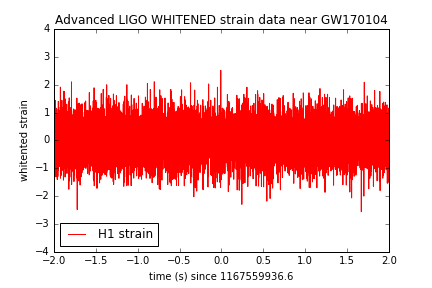}
\includegraphics[width=0.32\textwidth]{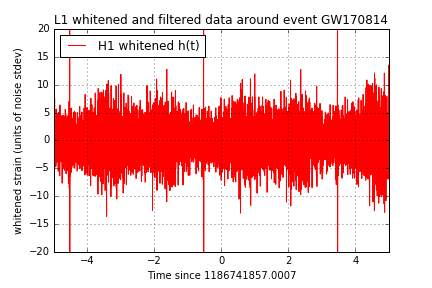}
\includegraphics[width=0.32\textwidth]{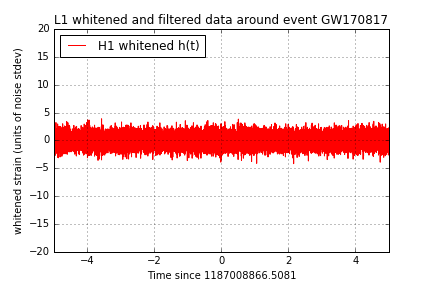}

\caption{Plots of H1 whitened and filtered strain in GW150914, GW151226, GW170104, GW170814 and GW170817.}
\label{fig:GW_amplitude}
\end{figure}

\begin{figure}[h!]
\centering
\includegraphics[width=0.96\textwidth]{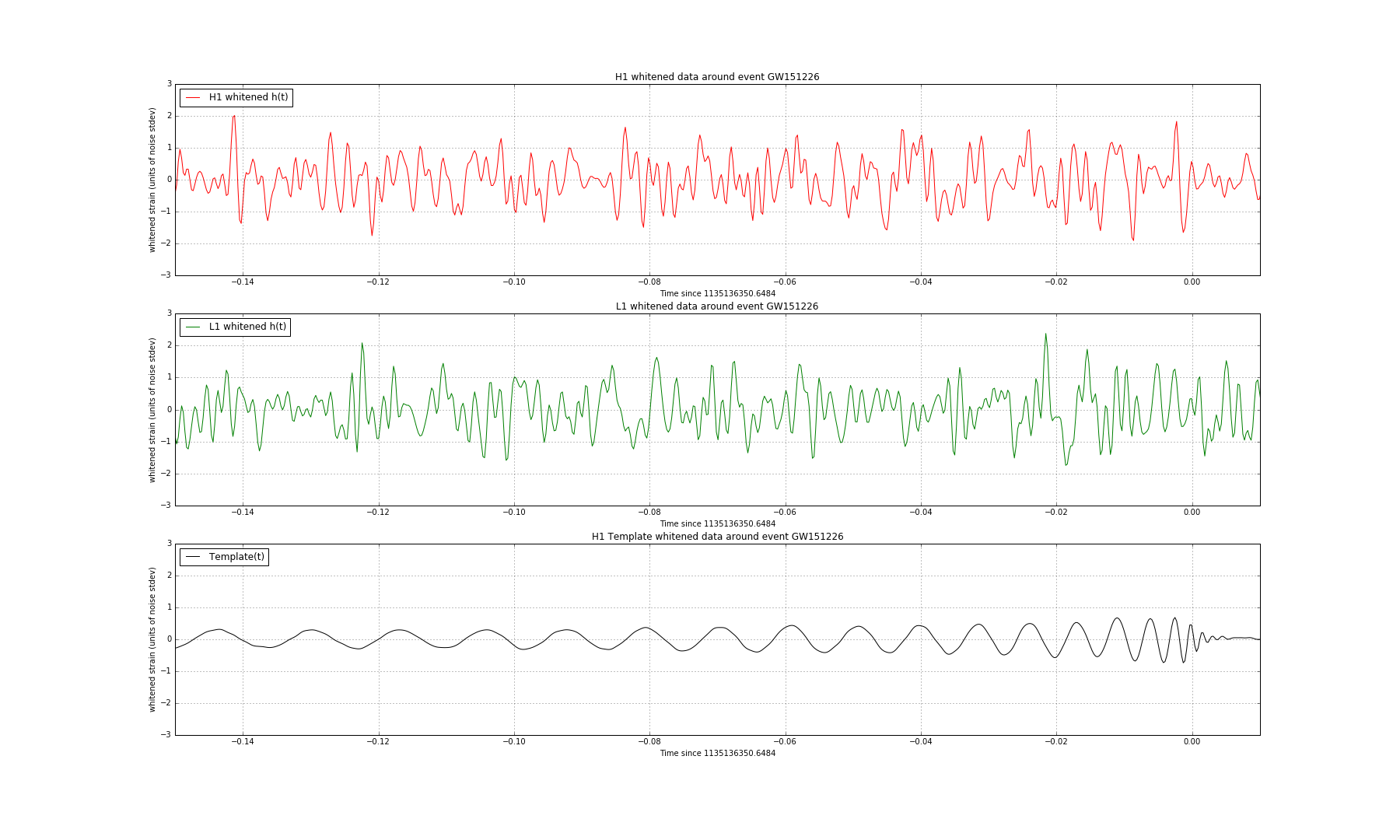}
\caption{GW151226 whitened and filtered H1 and L1 strain, and the template }
\label{fig:GW151226_signal}
\end{figure}

\begin{figure}[h!]
\centering
\includegraphics[width=0.96\textwidth]{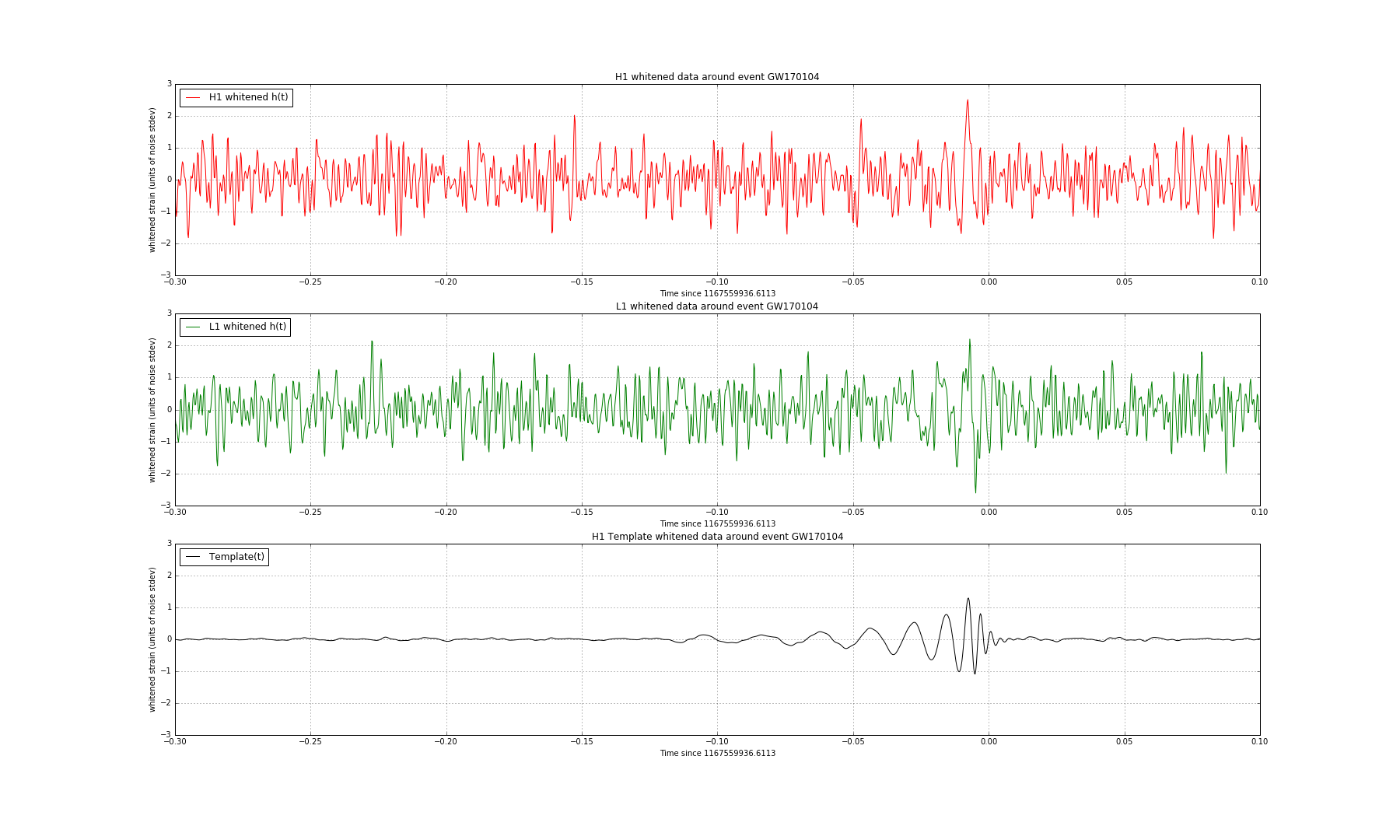}
\caption{GW170104 whitened and filtered H1 and L1 strain, and the template}
\label{fig:GW170104_signal}
\end{figure}

\begin{figure}[h!]
\centering
\includegraphics[width=0.96\textwidth]{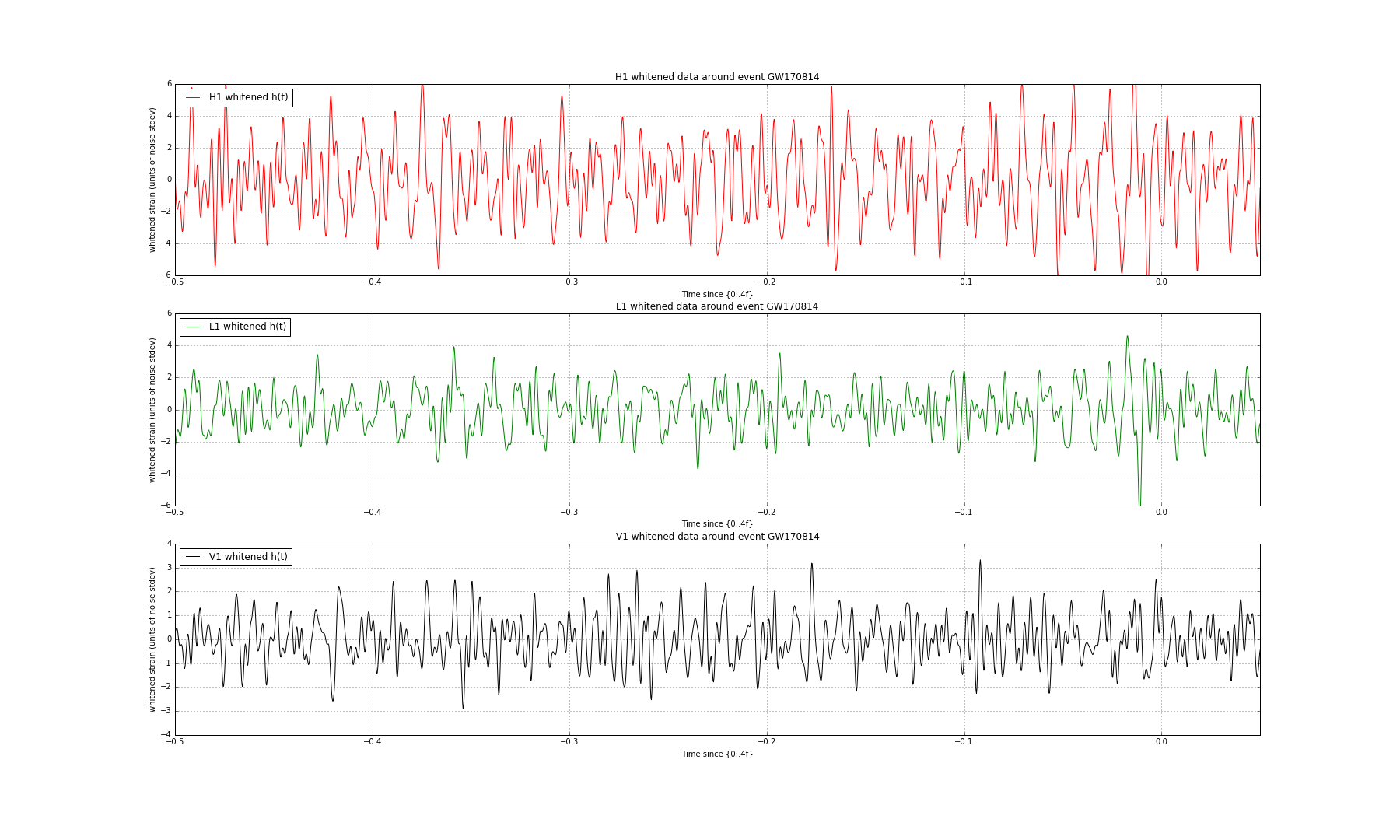}
\caption{GW170814 whitened and filtered H1, L1 , V1 strain}
\label{fig:GW170814_signal}
\end{figure}

\begin{figure}[h!]
\centering
\includegraphics[width=0.96\textwidth]{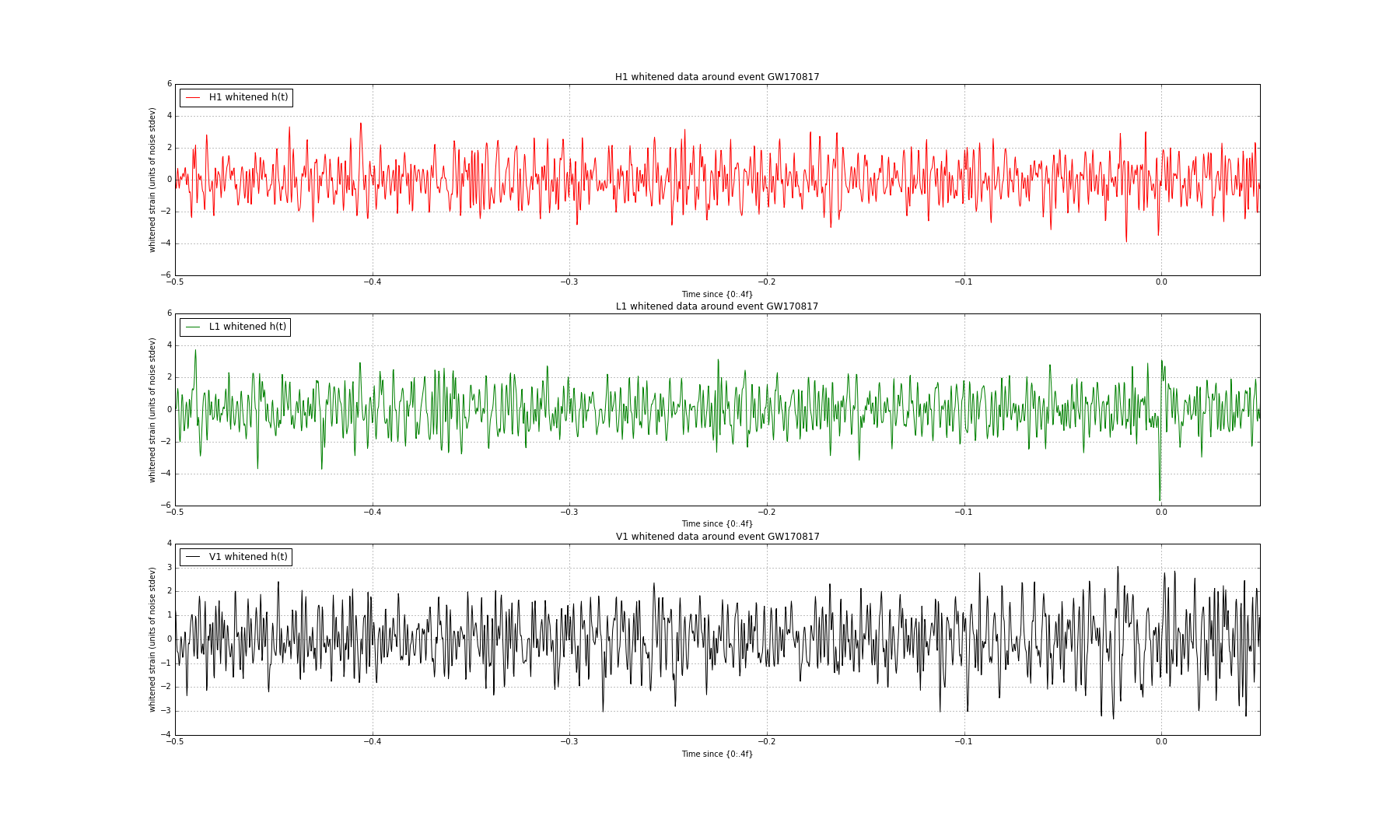}
\caption{GW170817 whitened and filtered H1, L1 , V1 strain}
\label{fig:GW170817_signal}
\end{figure}

\clearpage

\section*{Why False Alarm Rate calculation is unreliable}
\label{sec:false_alarm_rate}

GW151226  is observed over a duration of 1 second in the window $[t_1, t_2]$. Then LIGO software searches the vicinity for a duration of 10 days before or after the event, say $[t_a, t_b]$, where $t_a = t_1 - (11$ days) and $t_b = t_1 - (1$ day). It is possible that, during observation period $[t_a, t_b]$, there were no \textbf{coincident} noise bursts, and measured noise events $n_{b}(\hat{\rho_{c}})$ was low, and computed false alarm rate is 1 in 200,000 years. 

Given the non-stationarity of detector noise, it is possible that, \textbf{only} during the presumed GW event window $[t_1, t_2]$, there was indeed low amplitude noise burst, as in the top right panel in Fig.~\ref{fig:LIGO_matched_filter_sinewave}, causing high reweighted SNR in the matched filter for both H1 and L1 and high detection statistic value(DSV). This noise burst can be $\frac{1}{500}$ times \textbf{smaller} than the standard deviation of detector noise. 

There is simply \textbf{no way} we can make an argument that, just because LIGO team observed no coincident noise bursts in $[t_a, t_b]$, we cannot expect a coincident noise burst in $[t_1, t_2]$, because detector noise is non-stationary. Even a low-amplitude noise burst with standard deviation $\frac{1}{500}$ times \textbf{smaller} than the standard deviation of detector noise, can cause \textbf{coincident} high SNR in matched filter. It is also possible that, during observation period $[t_a, t_b]$, noise had \textbf{low mean} and did not cause high SNR frequently, while during GW event  window $[t_1, t_2]$,, non-stationary noise could have had \textbf{higher mean}, causing high SNR in the matched filter. We should \textbf{include} this event $[t_1, t_2]$, in  characterizing detector noise and \textbf{not pre-suppose} that these are valid GW events and then look in the vicinity only, \textbf{excluding} these events, for detector noise characterization and false alarm rate computation.

The false alarm rate computation associated with the $5 \sigma$ significance, using the expression $F(\hat{\rho_{c}}) = 1 - e^{\frac{-T}{T_b} (1+ n_{b}(\hat{\rho_{c}}))}$, is dependent on the critical parameter $n_{b}(\hat{\rho_{c}})$, which is empirically determined, by observations over a number of days, near the vicinity of the GW event. This empirical parameter $n_{b}(\hat{\rho_{c}})$ is the \textbf{weakest} link in the argument. The above counter-example with coincident low amplitude noise bursts, directly rebuts it. The  empirical parameter $n_{b}(\hat{\rho_{c}})$ has \textbf{no theoretical basis} and is\textbf{ not} a reliable predictor of false alarm rate. 

If the computed false alarm rate is 200,000 years, this only means that the \textbf{average time} between 2 successive false coincident detections is 200,000 years. This \textbf{does not mean} that, we need to wait for 200,000 years, following the observation period, to observe a false coincident detection. We could get the first false coincident detection a few months after the end of the observation period, and the second detection a few months later, as well. If we observe the detector noise over a period far larger than 200,000 years, we would expect coincident detections due to noise, with an \textbf{average interval} of 200,000 years.

This is the reason, normalized CCF of H1 vs L1 test is \textbf{crucial}, which will reject this \textbf{coincident} false detection. We can \textbf{rule out} this case of false coincident detection by the crucial new test of normalized CCF on H1 vs L1, explained in Section~\ref{sec:test-3b}.

\end{document}